\documentclass{JHEP3}

\usepackage{graphicx}

\usepackage{multicol}
\usepackage{bbm}
\usepackage{amsmath}
\usepackage{amssymb}
\usepackage{euscript}
\usepackage{array}
\usepackage{amsfonts}
\usepackage{mathrsfs}

\def\be{\begin{equation}}
\def\ee{\end{equation}}
\def\beq{\begin{equation}}
\def\eeq{\end{equation}}

\def\ba{\begin{eqnarray}}
\def\ea{\end{eqnarray}}

\title{CMB-Galaxy correlation in Unified Dark Matter Scalar Field Cosmologies}

\author{Daniele Bertacca$^{a,b,c}$, Alvise Raccanelli$^{c}$, Oliver F. Piattella$^{d,e}$ , Davide Pietrobon$^{f}$, Nicola Bartolo$^{a,b}$, Sabino Matarrese$^{a,b}$, Tommaso Giannantonio$^{g,h}$\\
$^a$ Dipartimento di Fisica Galileo Galilei, Universit\`{a} di Padova , via F. Marzolo, 8 I-35131 Padova, Italy\\
$^b$ INFN Sezione di Padova, via F. Marzolo, 8 I-35131 Padova, Italy\\
$^c$ Institute of Cosmology \& Gravitation, University of Portsmouth, Dennis Sciama Building, Portsmouth, PO1 3FX, United Kingdom\\
$^d$ Departamento de F\'{i}sica, Universidade Federal do Esp\'{i}rito Santo, avenida Ferrari 514, 29075-910 Vit\'{o}ria, ES, Brazil\\
$^e$ INFN, sezione di Milano, Via Celoria 16, 20133 Milano, Italy\\
$^f$ Jet Propulsion Laboratory, California Institute of Technology, 4800 Oak Grove Drive, 91109 Pasadena CA USA\\
$^g$ Excellence Cluster Universe, Technical University Munich, Boltzmannstra\ss e 2, D-85748 Garching bei M\"unchen, Germany\\
$^h$ Argelander--Institut f\"ur Astronomie der Universit\"at Bonn, Auf dem H\"ugel 71, D-53121 Bonn, Germany\\
E-mails: \email{daniele.bertacca@pd.infn.it}, \email{alvise.raccanelli@port.ac.uk}, \email{oliver.piattella@gmail.com}, \email{davide.pietrobon@jpl.nasa.gov}, \email{nicola.bartolo@pd.infn.it}, \email{sabino.matarrese@pd.infn.it}, \email{tommaso.giannantonio@Universe-cluster.de} }


\abstract{We present an analysis of the cross-correlation between the CMB and the large-scale structure (LSS) of the Universe in Unified Dark Matter (UDM) scalar field cosmologies. We work out the predicted cross-correlation function in UDM models, which depends on the speed of sound of the unified component, and compare it with observations from six galaxy catalogues (NVSS, HEAO, 2MASS, and SDSS main galaxies, luminous red galaxies, and quasars). We sample the value of the speed of sound and perform a likelihood analysis, finding that the UDM model is as likely as the $\Lambda$CDM, and is compatible with observations for a range of values of  $c_\infty$ (the value of the sound speed at late times) on which structure formation depends. In particular, we obtain an upper bound of $c_\infty^2 \le 0.009$ at 95\% confidence level, meaning that the $\Lambda$CDM model, for which $c_\infty^2 = 0$, is a good fit to the data, while the posterior probability distribution peaks at the value $c_\infty^2=10^{-4}$ .
Finally, we study the time dependence of the deviation from $\Lambda$CDM via a tomographic analysis using a mock redshift distribution and we find that the largest deviation is for low-redshift sources, suggesting that future low-z surveys will be best suited to constrain UDM models.}

\begin{document}

\section{Introduction}
In the last decade, observations of large-scale galaxy distribution \cite{2mass,sdss_dr6,Condon1998NVSS,heao}, the search for Type
Ia supernovae (SNIa) \cite{Perlmutter:1998np, Riess:1998cb, Riess:1998dv, Amanullah:2010vv} and measurements of the 
Cosmic Microwave Background anisotropies (CMB) \cite{Larson:2010gs, Komatsu:2010fb} have been suggesting that two unknown
components govern the dynamics of the Universe. They are referred to as dark matter (DM), necessary to explain the structure formation, and dark energy (DE), that is supposed to drive the measured cosmic acceleration (for a review see \cite{Tsujikawa:2010sc,Copeland:2006wr}). However, DM particles have
not been directly detected yet, although there are hints of their existence \cite{Adriani:2008zr, Adriani:2008zq, Bernabei:2000qi}, and there is no theoretically established motivation either for DE~\cite{Tsujikawa:2010sc,Copeland:2006wr} or for the tiny cosmological constant \cite{Weinberg:1988cp} which would fit cosmological observations \cite{Amendola:1272934}.

A key indication of an accelerated phase in the cosmic history is an excess of power in the low multipole region of the CMB angular power spectrum, likely a signature of the late integrated Sachs-Wolfe (ISW) effect \cite{Sachs:1967er}. The late ISW effect is caused by a time-evolving gravitational potential and it is interpreted as an effect of a new cosmological component which becomes the dominant contribution to the total energy content of the Universe at recent epochs. For this reason, it would affect large cosmological scales. Cross-correlating the distribution of galaxies with the CMB \cite{Crittenden:1995ak} has been proven to increase the signal-to-noise ratio \cite{Boughn:2001zs, Boughn:2003yz} and to be a useful probe of the late-time evolution of the Universe. 

The first correlation between WMAP \cite{Jarosik:2010} data and the HEAO-1 X-ray map \cite{Boldt:1987} was measured by \cite{Boughn:2001zs, Boughn:2003yz}  and then confirmed by the WMAP team \cite{Nolta:2003uy} using a radio sources map from the NRAO VLA Sky Survey \cite{Condon1998NVSS}. Further analyses based on different and complementary techniques strengthened the evidence for the ISW effect \cite{Vielva:2006,Cabre:2006qm, Gaztanaga:2004sk,Giannantonio:2006,Pietrobon:2006,McEwen2007,Raccanelli:2008kx,Ho:2008bz,Giannantonio:2008,Granett:2008ju,Xia:2010pe} (see \cite{Aghanim:2007bt} for a review) and allowed to test several aspects of cosmological models, focusing on properties of DE \cite{Hu:2004yd, Corasaniti:2005pq,Giannantonio:2006ij, Xia:2009,Xia:2009dr}, 
evolution of radio galaxies and their bias \cite{Massardi:2010sx} and alternatives to general relativity such as DGP \cite{Giannantonio:2008fk} and Galileon models (see for example \cite{Nesseris:2010pc, DeFelice:2010pv, Tsujikawa:2010sc}); this technique could in future be applied to even broader areas, such as cosmic reionization \cite{Giannantonio:2007za}.

The standard cosmological model $\Lambda$CDM is a good fit to the current observations, but it describes the Universe by means of two unknown components which represent the 95\% of the total energy density. This is theoretically unsatisfactory and alternative models which describe DM and DE from an unique prospective have been proposed, such as a single component behaving both as dark matter and dark energy, which has been often 
referred to in literature as ``Unified Dark Matter" (UDM), or ``Quartessence",  see e.~g.~\cite{Kamenshchik:2001cp,Bilic:2001cg,Bento:2002ps,Carturan:2002si, Amendola:2003bz, Sandvik:2002jz,Makler:2003iw, Scherrer:2004au, Giannakis:2005kr, Bertacca:2007ux, Bertacca:2007cv, Bertacca:2007fc, Quercellini:2007ht, Balbi:2007mz, Bertacca:2008uf,Pietrobon:2008js, Bilic:2008yr, Camera:2009uz, Li:2009mf, Chimento:2009nj, Piattella:2009kt, Gao:2009me, Camera:2010wm, Lim:2010yk, Bertacca-2010-2, transfer-function}.
See \cite{Bertacca:2010ct} for a recent review with an up-to-date list of UDM models. 
In comparison with the standard DM + DE models (e.g.\ even the
simplest case, with DM  and a cosmological constant), these models have 
the advantage that they can describe the dynamics of the Universe with a single scalar field which
triggers both the accelerated expansion at late times and the structure formation at earlier times. 
Specifically, for these models, we can use Lagrangians with a non-canonical kinetic term (see, for example,  \cite{Mukhanov:2005sc}), 
namely a Lagrangian which is an arbitrary function of the scalar field and of the kinetic term of the scalar field. In \cite{Bertacca:2008uf}, 
the authors proposed and studied a technique to reconstruct models where the
effective speed of sound of the scalar field is small enough that the scalar field can cluster. These
models avoid the strong time evolution of the gravitational
potential and the large ISW effect \cite{Bertacca:2007cv}
which have been a serious drawback of previously considered
models. In this reconstruction technique first of all the scalar
field Lagrangian was required to be constant along the classical
trajectories. Specifically, by demanding that
$\mathcal{L}=-\Lambda$ on cosmological scales, the background evolution is
identical to that of the $\Lambda$CDM \cite{Bertacca:2007ux}. Secondly, from this result, they deduced that the energy-momentum tensor of this scalar field is made by two components: one
behaving like a pressure-less fluid, and the other having negative
pressure.

In this paper we consider the cosmological models of Unified Dark Matter introduced in \cite{Bertacca:2008uf} 
focusing on their predictions for the ISW-Large Scale Structure (LSS) cross-correlation. We compare them to current observations and constrain the sound speed of the scalar field, which is the characterising parameter of the model.

The paper is organized as follows. In Section~\ref{Int-UDM} we recall the basics of UDM models, specifying the Lagrangian and the parametric form of the sound speed for the model we consider and its implications for the gravitational potential evolution. In Section~\ref{3} we describe the theoretical cross-correlation functions for the large scale surveys discussed in \cite{Giannantonio:2008} (NVSS, SDSS main galaxies, LRGs and quasars, HEAO, 2MASS). In Sections~\ref{sec:data_analysis}-\ref{results} we perform the data analysis and discuss the results.  Finally in Section~\ref{conclusions} we draw our conclusions.

\section{Unified Dark Matter Scalar field models}
\label{Int-UDM}

We start recalling the main equations which are useful for the description of most the UDM models within the framework of k-essence  (see, for example,  \cite{Mukhanov:2005sc}). 
Let us consider the following action
\begin{equation}\label{eq:action}
S = S_{G} + S_{\varphi} =  \int d^4 x \sqrt{-g} \left[\frac{R}{2}+\mathcal{L}(\varphi, X)\right]\;,
\end{equation}
where
\begin{equation}\label{x}
X = -\frac{1}{2}\nabla_\mu \varphi \nabla^\mu \varphi\;.
\end{equation}
We adopt $8\pi G = c^2 = 1$ units and the $(-,+,+,+)$ signature for the metric (Greek indices run over space-time dimensions, while Latin indices label spatial coordinates).

The energy-momentum tensor of the scalar field $\varphi$ has the following form
\begin{equation}\label{energy-momentum-tensor}
 T^{\varphi}_{\mu \nu } = - \frac{2}{\sqrt{-g}}\frac{\delta S_{\varphi }}{\delta g^{\mu \nu }}=\frac{\partial \mathcal{L}(\varphi ,X)}{\partial X}\nabla_{\mu }\varphi\nabla _{\nu }\varphi +\mathcal{L}(\varphi ,X)g_{\mu \nu }\;.
\end{equation}
If $X$ is time-like, $S_{\varphi}$ describes a perfect fluid and the energy-stress tensor becomes $T^{\varphi}_{\mu \nu }= (\rho + p)u_{\mu}u_{\nu } + p\,g_{\mu \nu }$ which is characterised by the fluid density, $\rho,$ and pressure, $p$, once the following identifications are made
\begin{eqnarray}
  \label{pressure}
&& p(\varphi ,X) = \mathcal{L}(\varphi, X)\;, \\
\label{energy-density}
&&  \rho(\varphi ,X)= 2X\frac{\partial p(\varphi ,X)}{\partial X}-p(\varphi ,X)\;,
\end{eqnarray}
 and the four-velocity is defined as
\begin{equation}
  \label{eq:four-velocity}
  u_{\mu }= \frac{\nabla _{\mu }\varphi }{\sqrt{2X}}\;.
\end{equation}
Moreover, let us introduce the equation of state parameter (EoS), defined as the pressure-to-density ratio: $w\equiv p/\rho$.

We will use a flat Friedmann-Lema\^{i}tre-Robertson-Walker (FLRW) metric $ds^2=-dt^2+a(t)^2\delta_{ij} dx^i dx^j = a(\eta)^2 (-d\eta^2 + \delta_{ij} dx^i dx^j) $, 
where $a(t)$ is the scale factor and $\eta$ is the conformal time, and we assume 
that the energy density of the radiation is negligible at the times of interest. Disregarding also the small baryonic component, 
 the background evolution of the Universe is completely characterised by the following equations:
\begin{eqnarray}\label{eq_u1}
&&\mathcal{H}^2 = a^2 H^2 = \frac{1}{3} a^2 \rho\;, \\
\label{eq_u2}
&&\mathcal{H}'-\mathcal{H}^2=a^2 \dot{H} = - \frac{1}{2} a^2(p + \rho)\;,
\end{eqnarray}
where ${\cal H} = a'/a$ and $H = \dot{a}/a$, the dot denoting differentiation w.r.t. the cosmic time $t$ whereas a prime w.r.t. the conformal time $\eta$.

In the background we have that $X = \dot{\varphi}^2/2=\varphi'^2/(2a^2)$, therefore the equation of motion for the homogeneous mode $\varphi(t)$ becomes
\begin{equation}\label{eq_phi}
\left(\frac{\partial p}{\partial X} + 2X\frac{\partial^2 p}{\partial X^2}\right)\ddot\varphi + \frac{\partial p}{\partial X}(3H\dot\varphi) + \frac{\partial^2 p}{\partial \varphi \partial X}\dot\varphi^2 - \frac{\partial p}{\partial \varphi} = 0\;.
\end{equation}
To the first order in the (scalar) perturbations, we consider small inhomogeneities of the scalar field $\varphi(t,\mbox{\boldmath x})= \varphi_0(t)+\delta\varphi(t,\mbox{\boldmath $x$})$ and 
we expand the FLRW metric in the longitudinal gauge as 
\begin{equation}
ds^2 = -(1 + 2 \Phi)dt^2 + (1 - 2 \Phi)a(t)^2\delta_{ij} dx^i dx^j\;, 
\end{equation}
being $\delta T_{i}^j = 0$ for $i \neq j$ \cite{Mukhanov:1990me}.

The first-order part of $(0-0)$ and $(0-i)$ component of the energy-stress tensor are
(see Ref.~\cite{Garriga:1999vw} and Ref.~\cite{Mukhanov:2005sc})
\begin{eqnarray}
\delta T^{\varphi \;0}_{\phantom{\varphi \;}0}&=&-\delta \rho=-\frac{\partial  \rho}{\partial \phi} \delta\phi 
- \frac{\partial  \rho}{\partial X}  \delta X= \nonumber \\
&=&-\frac{p+\rho}{c_{\rm s}^2}\left[\left(\frac{\delta\varphi}{\varphi_0'} \right)'
+\mathcal{H} \frac{\delta\varphi}{\varphi_0'}- \Phi\right]+
3\mathcal{H}(p+\rho) \frac{\delta\varphi}{\varphi_0'}\;, \\
\delta T^{\varphi \;0}_{\phantom{\varphi \;}i}&=&-(p+\rho)\left( \frac{\delta\varphi}{\varphi_0'} \right)_{,i}\;,
\end{eqnarray}
where one defines a ``speed of sound'' $c_{\rm s}^2$ as
\begin{equation}\label{cs}
c_{\rm s}^2 \equiv  \frac{\partial p /\partial X}{\partial \rho /\partial X} = \frac{\partial p/\partial X}{(\partial p/\partial X)+ 2X(\partial^2p/\partial X^2)} \;. 
\end{equation}
Finally, linearizing the Einstein equations one obtains \cite{Garriga:1999vw, Mukhanov:2005sc}
\begin{equation}\label{pertu-eq1}
\nabla^2 \Phi = \frac{1}{2} \frac{a^2(p+\rho)}{c_{\rm s}^2 \mathcal{H}} \left(\mathcal{H}  \frac{\delta\varphi}{\varphi_0'}+ \Phi \right)'\;,
\end{equation}
and
\begin{equation}
\label{pertu-eq2}
\left(a^2 \frac{\Phi}{\mathcal{H}} \right)'= \frac{1}{2} \frac{a^2(p+\rho)}
{\mathcal{H}^2}\left(\mathcal{H}  \frac{\delta\varphi}{\varphi_0'}
+ \Phi \right) \;.
\end{equation}
Eqs.~(\ref{pertu-eq1}) and~(\ref{pertu-eq2}) are sufficient to determine the gravitational potential $\Phi$ and the perturbation of the scalar field. It is useful to write explicitly the perturbed scalar field as a function of the gravitational potential
\begin{equation}
\label{delta-phi}
 \frac{\delta\varphi}{\varphi_0'}=2\frac{\Phi'+ \mathcal{H}\Phi}{a^2(p+\rho)}\;.
\end{equation}
Defining two new variables
\begin{equation}
\label{u-v}
u\equiv 2 \frac{\Phi}{(p+\rho)^{1/2}} \;, \quad\quad v\equiv z \left(\mathcal{H}  \frac{\delta\varphi}{\varphi_0'}+ \Phi \right)\;,
\end{equation}
where $z = a^2(p+\rho)^{1/2}/(c_{\rm s} \mathcal{H})$, we can recast (\ref{pertu-eq1}) and (\ref{pertu-eq2}) in terms of $u$ and $v$ \cite{Mukhanov:2005sc}:
\begin{equation}
\label{pertu-eq_uv}
c_{\rm s} \nabla^2 u = z \left( \frac{v}{z}\right)'\;,\quad \quad 
c_{\rm s} v= \theta \left( \frac{u}{\theta}\right)'
\end{equation}
where $\theta = 1/(c_{\rm s} z)=(1+p/\rho)^{-1/2}/(\sqrt{3}a)$. Starting from (\ref{pertu-eq_uv}) we arrive at the following second order differential equations for $u$ \cite{Mukhanov:2005sc}:
\begin{equation}\label{diff-eq_u}
u''- c_{\rm s}^2 \nabla^2 u - \frac{\theta''}{\theta}u = 0\;.
\end{equation}
Unfortunately, we do not know the exact solution for a generic Lagrangian. However, we can consider the asymptotic solutions, i.e. the long-wavelength and the short-wavelength perturbations,  corresponding to the regimes $c_{\rm s}^2k^2 \ll \left|\theta''/\theta\right|$  and $c_{\rm s}^2k^2 \gg \left|\theta''/\theta\right|$, respectively. This amounts to consider perturbations on scales much larger or much smaller than the effective comoving Jeans length for the gravitational potential \cite{Bertacca:2007cv}, which is defined as follows:
\begin{equation}
\label{Jeans}
\lambda^2_{\rm J}(\eta) = 2\pi/ k_{\rm J}^2 \equiv 2\pi c_{\rm s}^2 \left|\theta/\theta''\right|\;,
\end{equation}
where $k_{\rm J}^2$ is the squared Jeans wave number \cite{Bertacca:2007cv}. For a plane wave perturbation $u \propto u_{\bf k} (\eta) \exp(i \mbox{\boldmath $k \cdot x$})$ in the short-wavelength limit ($c_{\rm s}^2k^2 \gg \left|\theta''/\theta\right|$) we obtain from Eq.~\eqref{diff-eq_u}
\begin{equation}
\label{u-cs_k>>theta''/theta}
u_{\bf k} \simeq\frac{C_{\bf k}(\bar{\eta})}{c_{\rm s}^{1/2}(\eta)}
\cos \left( k \int_{\bar{\eta}}^{\eta} c_{\rm s} d\tilde{\eta}\right)\;,
\end{equation}
where $C_{\bf k}$ is a constant of integration at some initial time $\bar{\eta}$. 
In the opposite regime, neglecting the decaying mode, the long-wavelength solution ($c_s^2k^2 \ll \left|\theta''/\theta\right|$) is
\begin{equation}
\label{u-cs_k<<theta''/theta}
u_{\bf k} = A_{\bf k}(\bar{\eta}) \theta \int_{\bar{\eta}}^{\eta}\frac{d\tilde{\eta}}{\theta^2}\;,
\end{equation}
where $A_{\bf k}$ is a constant of integration.

\subsection{Matter density and Gravitational potential}
The authors of \cite{Bertacca:2008uf} proposed a technique to construct UDM models where the scalar field can have a sound speed small enough to allow for structure formation and to avoid a strong integrated Sachs-Wolfe effect \cite{Sachs:1967er} in the CMB anisotropies which typically plagues UDM models \cite{Carturan:2002si, Amendola:2003bz, Bertacca:2007cv} (see also \cite{Camera:2009uz,Camera:2010wm}). By applying this technique we can explicitly derive the functional forms of the energy density and gravitational potential for the class of model we are interested in.

\subsubsection{Energy density}
In particular, starting from the following scalar field Lagrangian $\mathcal{L}$ \cite{Bertacca:2008uf}
\begin{eqnarray}
\mathcal{L}(\varphi,X)&=&-\frac{\Lambda c_\infty}{(1- c_\infty^2)}
\frac{\cosh\left(\gamma\varphi\right)}{\sinh\left(\gamma\varphi\right)
\left[1+(1-c_\infty^2)\sinh^2\left(\gamma \varphi\right)\right]} \sqrt{1-\frac{2X}{\Lambda}}\nonumber\\ 
&&- \frac{\Lambda}{(1-c_\infty^2)}\frac{\left[(1-c_\infty^2)^2\sinh^2
\left(\gamma\varphi\right)+1-c_\infty^2\right]}{1+(1-c_\infty^2)\sinh^2
\left(\varphi\right)}\;,
\end{eqnarray}
where $\gamma=\left[3/[4(1-c_\infty^2)]\right]^{1/2}$, and once the initial value of $\varphi$ is fixed at early times, the scalar field Lagrangian is constant and equal to $-\Lambda$ along the classical trajectories. Specifically, when $\mathcal{L} =  p = -\Lambda$ on cosmological scales, the background is identical to the one of the $\Lambda$CDM. Indeed, if we consider the equation of motion of the scalar field, and if $p = -\Lambda$, we easily obtain
\begin{equation}
\rho\left[a(t)\right]=\rho_\mathrm{DM}(a=1)a^{-3}+\Lambda\equiv\rho_\mathrm{DM}+\rho_{\Lambda }\, ,
\end{equation}
where $\rho_\Lambda$ behaves like a cosmological constant ($\rho_\Lambda=\mathrm{const.}$) and $\rho_\mathrm{DM}$ behaves like a DM component ($\rho_\mathrm{DM}\propto a^{-3}$). This result implies that we can think of the stress-tensor of the scalar field as being made of two components: one behaving like a pressureless fluid, and the other having negative pressure. Therefore the integration constant $\rho_\mathrm{DM}(a = 1)$ can be interpreted as the ``dark matter'' component today with $\Omega_{\rm m 0}=\rho_\mathrm{DM}(a=1)/(3{H_0}^2)$ and $\Omega_{\Lambda 0}=\rho_\Lambda/(3{H_0}^2)$ the density parameters of DM and DE. In this case, the parametric form for the sound speed is \cite{Bertacca:2008uf}
\begin{equation}
\label{eq:cs2}
{c_{\rm s}}^2(a)=\frac{{\Omega_{\Lambda 0} c_\infty}^2}{\Omega_{\Lambda 0}+(1-{c_\infty}^2)\Omega_{\rm m 0}a^{-3}}\;,
\end{equation}
where $c_\infty$ is the value of the sound speed when $a\rightarrow\infty$. Let us emphasise that when $a\to0$, then $c_{\rm s}\to0$.

Specifically, our class of UDM models allows the value $w=-1$ for $a\to\infty$. In other words, they admit an effective cosmological constant energy density at late times. Therefore, in order to compare the predictions of our UDM model with observational data, we follow similarly the prescription used in \cite{Piattella:2009kt}, where the density contrast of the clustering fluid is $\delta_{\rm DM}\equiv \delta \rho/\rho_{\rm DM }$ and where $\rho_{\rm DM}=\rho -\rho_\Lambda$, by definition, is  the only component of the scalar field density which clusters (for adiabatic EoS see \cite{Pietrobon:2008js, Piattella:2009kt}).  According to this notation we can derive the expression for the density contrast as function of the gravitational potential for scales smaller than the cosmological horizon and $z < z_{\rm rec}$, where $z_{\rm rec}$ is the recombination redshift ($z_{\rm rec} \approx 10^{3}$):
\begin{equation}
\label{delta-dm}
 \delta_{\rm {DM}}\left({\bf k};z\right) = \frac{\delta\rho\left({\bf k};z\right)}{3H_0^2\Omega_{\rm m0}\left(1 + z\right)^3} = -\frac{k^{2} \Phi \left({\bf k};z\right) }{\left(3/2\right)H_0^2\Omega_{\rm m0}\left(1 + z\right)}\;.
\end{equation}

\subsubsection{Gravitational potential}
Let us stress here an important point which holds for the models that we are considering (i.e. those which reproduce the same expansion history of the $\Lambda$CDM Universe): the gravitational potential evolves in the same way as in a $\Lambda$CDM Universe for those modes with wavelengths larger than the sound horizon. In fact, we can write the usual solution \cite{Hu:1998tj, Dodelson:2003ft}
\begin{equation}
\label{Phi_k-eta<1}
\Phi_{k\ll 1/\lambda_{\rm J}}({\bf k};z)=  B_{\bf k} \left(1-\frac{\mathcal{H}(\eta)}{a^2(\eta)} \int_{\eta_i}^{\eta}
a^2(\tilde{\eta})d\tilde{\eta}\right)\;, 
\end{equation}
for $k^2 \ll 1/\lambda^2_{\rm J}(\eta)$; here $B_{\bf k} \simeq (9/10)\Phi_{\rm p}({\bf k})T_{\rm m}(k)$, where $\Phi_{\rm p}({\bf k})$ is the primordial gravitational potential at large scales, set during inflation, see e.g.\ \cite{Dodelson:2003ft}, and 
$T_{\rm m}(k)$ is the matter transfer function suggested in \cite{Eisenstein:1997ik}, that includes baryons (a similar approach is been used also in Ref.\ \cite{Camera:2010wm}).

In this regime, $k^2 \ll 1/\lambda^2_{\rm J}(\eta)$, we can write $\Phi({\bf k};z)\simeq (9/10) \Phi_{\rm p}({\bf k})T_{\rm m}(k) (1+z) D(z)$, where $D(z)=g(z)/(1+z)$ is the growth factor and $g(z)$ the growth suppression factor for a $\Lambda$CDM Universe. A 
very good approximation for $g(z)$ as a function of redshift $z$ is given in Refs.~\cite{Lahav:1991wc,Carroll:1991mt, Eisenstein:1997ij}:
\begin{eqnarray}
g(z) =  \frac{5}{2}\Omega_m(z) \left\{ \Omega_m(z)^{4/7} -
\Omega_{\Lambda}(z) + \left[ 1+ \frac{\Omega_m(z)}{2} \right] \right. \left. \left[1+ \frac{\Omega_{\Lambda}(z)}{70} \right] \right\}^{-1}\;,
\end{eqnarray}
 with $\Omega_{\rm m}=\Omega_{\rm m0}(1+z)^3/E^2(z)$, $\Omega_\Lambda=\Omega_{\Lambda 0}/E^2(z)$, $E(z) \equiv (1+z) {\mathcal H}(z)/{\mathcal H}_0 = \left[\Omega_{\rm m0}(1+z)^3 + \Omega_{\Lambda0}\right]^{1/2}$ and $\Omega_{m0}$, $\Omega_{\Lambda0}=1-\Omega_{m0}$, the present-day density parameters of non-relativistic matter and cosmological constant, respectively. We have normalised the growth suppression factor so that $g=1$ early on when $\Omega_{\rm m} \rightarrow 1$ and $\Omega_\Lambda \rightarrow 0$. 

\noindent
In the opposite regime we have
\begin{equation}
\label{Phi_k-eta>1}
\Phi_{k\gg 1/\lambda_{\rm J}}({\bf k};\eta)=\frac{1}{2}\left[\frac{p+\rho}{c_{\rm s}}\right]^{1/2}(\eta)
 \mathit{C}_{\bf k}(\bar{\eta}) \cos\left(k\int_{\bar{\eta}}^{\eta}c_s(\tilde{\eta})d\tilde{\eta} \right)\quad , \quad 
\textrm{when} \quad k^2 \gg 1/\lambda^2_{\rm J}(\eta)\, .
\end{equation}
In Eq.~(\ref{Phi_k-eta>1}) we have $\mathit{C}_{\bf k}(\bar{\eta})  \simeq (9/10)\Phi_{\rm p}({\bf k})T_{\rm m}(k)\bar{\mathit{C}}$, where $\bar{\mathit{C}}$ is a constant of integration
\begin{equation}
\label{Cbar}
\bar{\mathit{C}}=2\frac{\left(1-\frac{\mathcal{H}(\bar{\eta})}{a^2(\bar{\eta})}
\int_{\eta_i}^{\bar{\eta}}a^2(\tilde{\eta})d\tilde{\eta}
\right)}{\left[(p+\rho)/c_{\rm s}\right]^{1/2}(\bar{\eta})}\; .
\end{equation}  
The value of $\bar{\mathit{C}}$ is obtained under the approximation that for $\eta <\bar{\eta}$ one can use the long wavelength solution~(\ref{Phi_k-eta<1}). In other words, we are assuming that there is an epoch in the past when the sound speed is very close to zero (a condition that is necessary in unified models to allow for structure formation). Notice that Eq.~(\ref{Phi_k-eta>1}) clearly shows that the gravitational potential is oscillating and decaying in time for the perturbations inside the sound horizon. Moreover, through Eqs.~(\ref{delta-dm}),  we can draw the same conclusion also for $\delta_{\rm DM}({\bf k},z)$.

In order to connect these two regimes, we can introduce a scale-dependent growth factor $D(k;z)$ as $\Phi({\bf k};z)\simeq (9/10) \Phi_{\rm p}({\bf k})T_{\rm m}(k) (1+z) D(k; z)$. In the 
long-wavelength regime, $k^2 \ll 1/\lambda^2_{\rm J}(\eta)$, $D(k;z)$ just reduces to $D(z)=g(z)/(1+z)$, according to the previous discussion. In this way one can naturally define a transfer 
function, $T_{\rm UDM}$, as 
\begin{eqnarray}
D(k; z)  &=& T_{\rm UDM}(k; \eta(z))D(z)\;, 
\end{eqnarray}
or, in other words, 
\begin{eqnarray}
\label{PhiT}
\Phi({\bf k};\eta) &=&T_{\rm UDM}(k; \eta)\Phi_{k\ll 1/\lambda_{\rm J}}({\bf k};\eta)\;.
\end{eqnarray}
In particular, we impose that $T_{\rm UDM}(k,\eta)=1$ for $\eta<\eta_{\rm rec}$, where $\eta_{\rm rec}$ is some epoch when the Universe is matter dominated and the radiation is negligible (usually later than the recombination epoch). Before this epoch the sound speed must be very close to zero to allow for structure formation.

\begin{figure}[h!]
\begin{center}
\includegraphics[width = 0.8\textwidth]{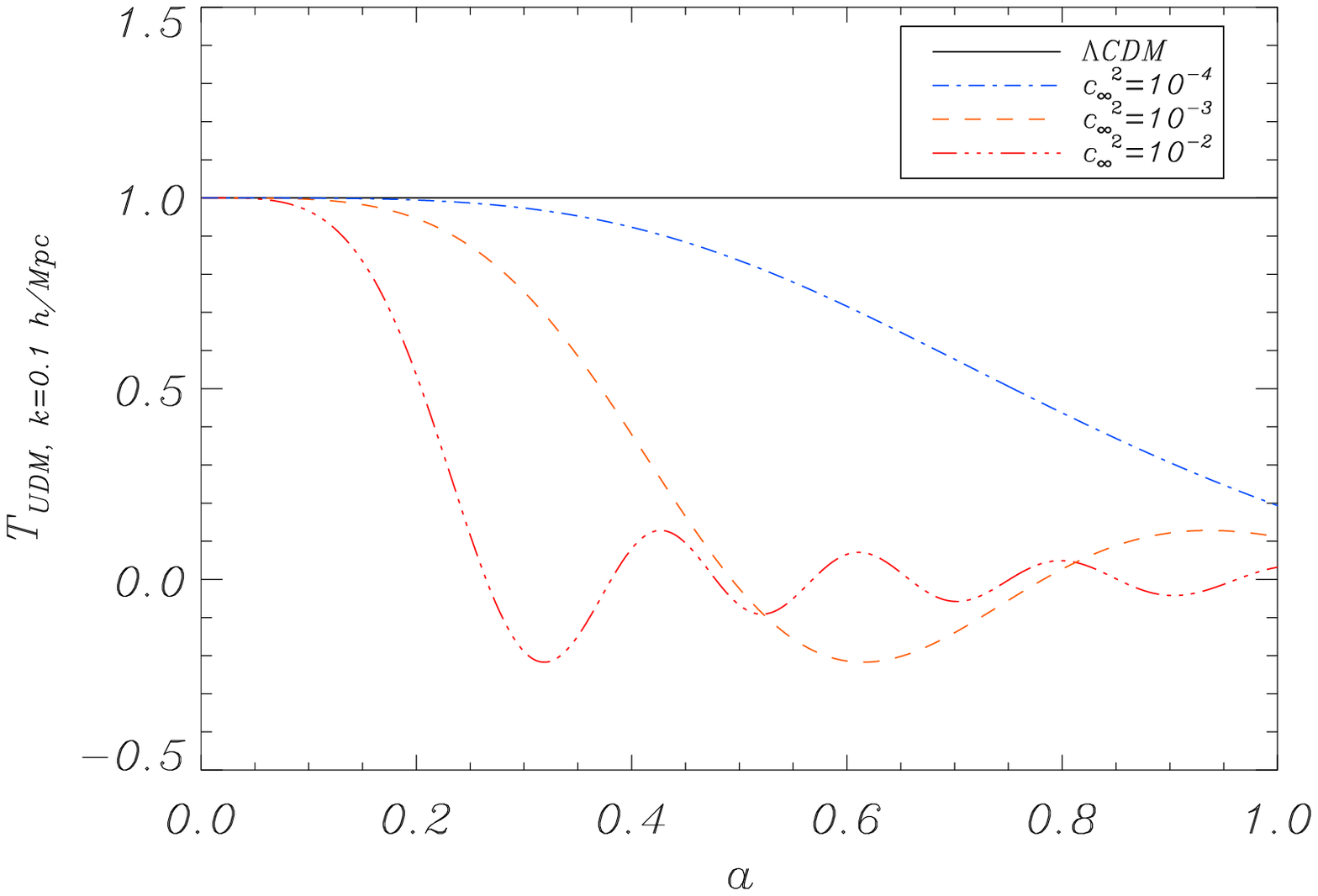}
\includegraphics[width = 0.8\textwidth]{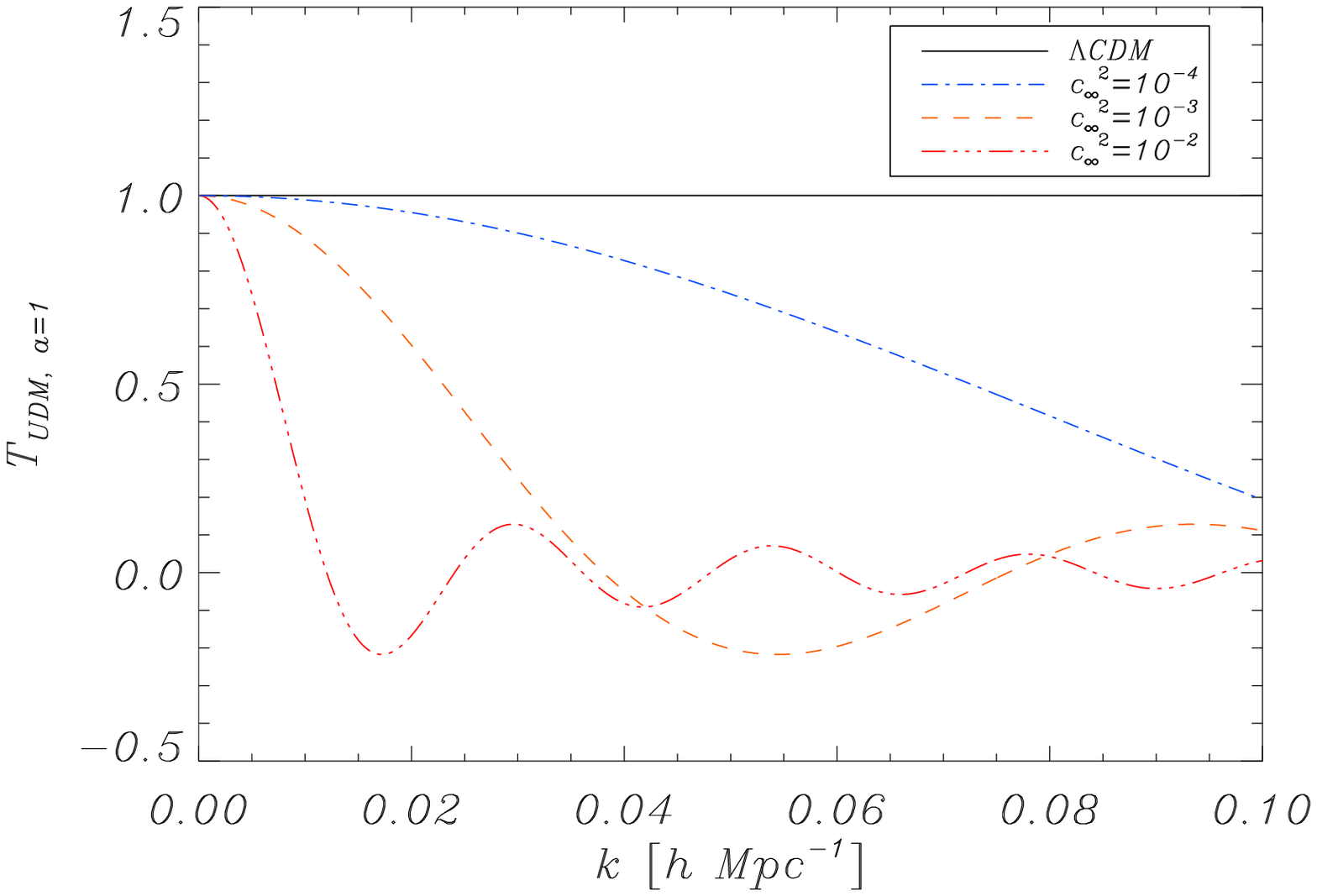}
\caption{Illustrative plot of $T_{\rm UDM}$ as a function of the scale factor for $k = 0.1 \;h$ Mpc$^{-1}$ (top panel) and as a function of the wavenumber for $a=1$ (bottom panel). 
For each panel, from left to right, the curves correspond to the choice $c_\infty^2 = 10^{-2}, 10^{-3}, 10^{-4}$ and $\Lambda$CDM model.}
\label{fig:A1}
\end{center}
\end{figure}

\begin{figure}[!ht]
\begin{center}
\includegraphics[width = .45\textwidth]{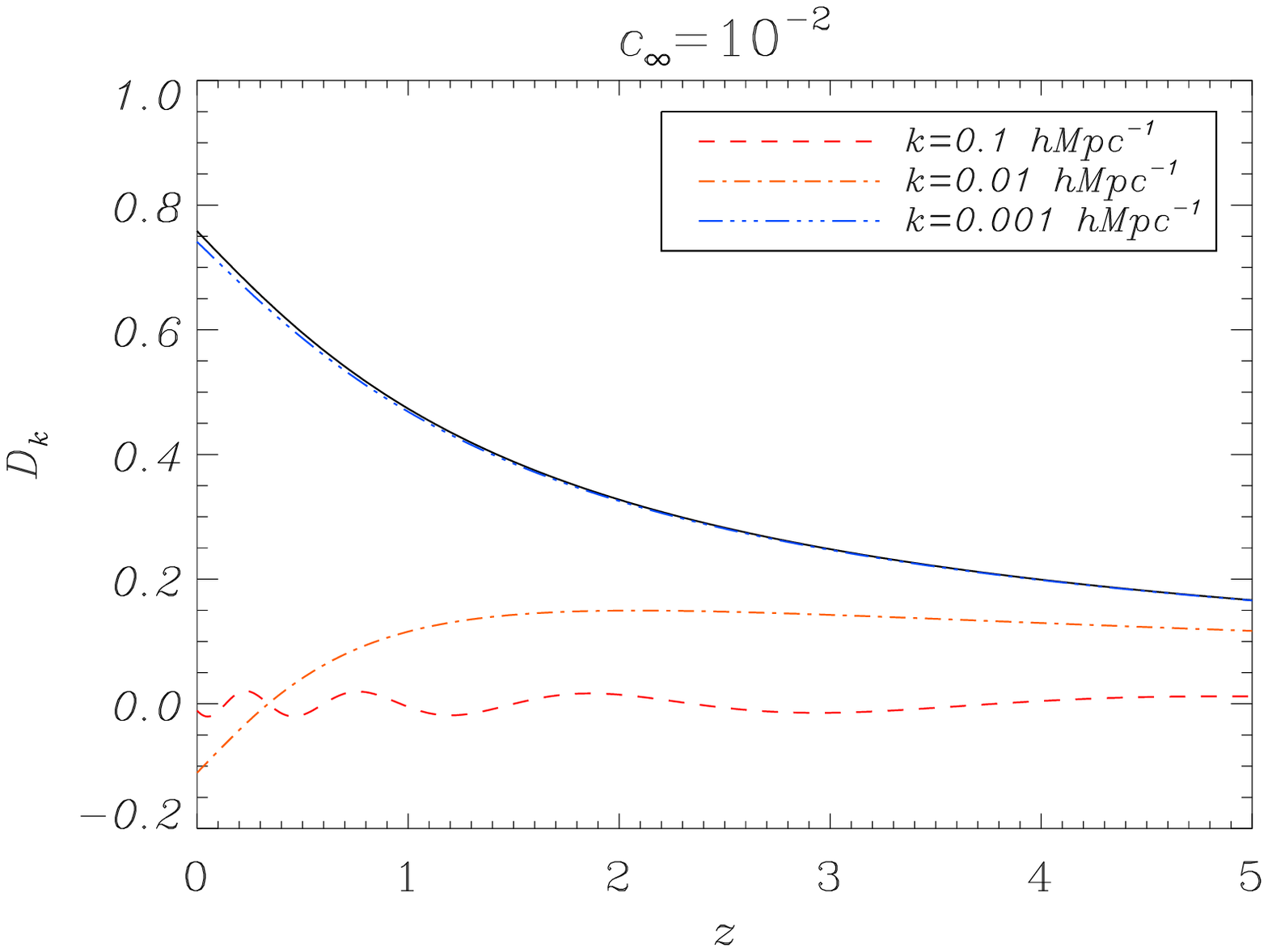}
\includegraphics[width = .45\textwidth]{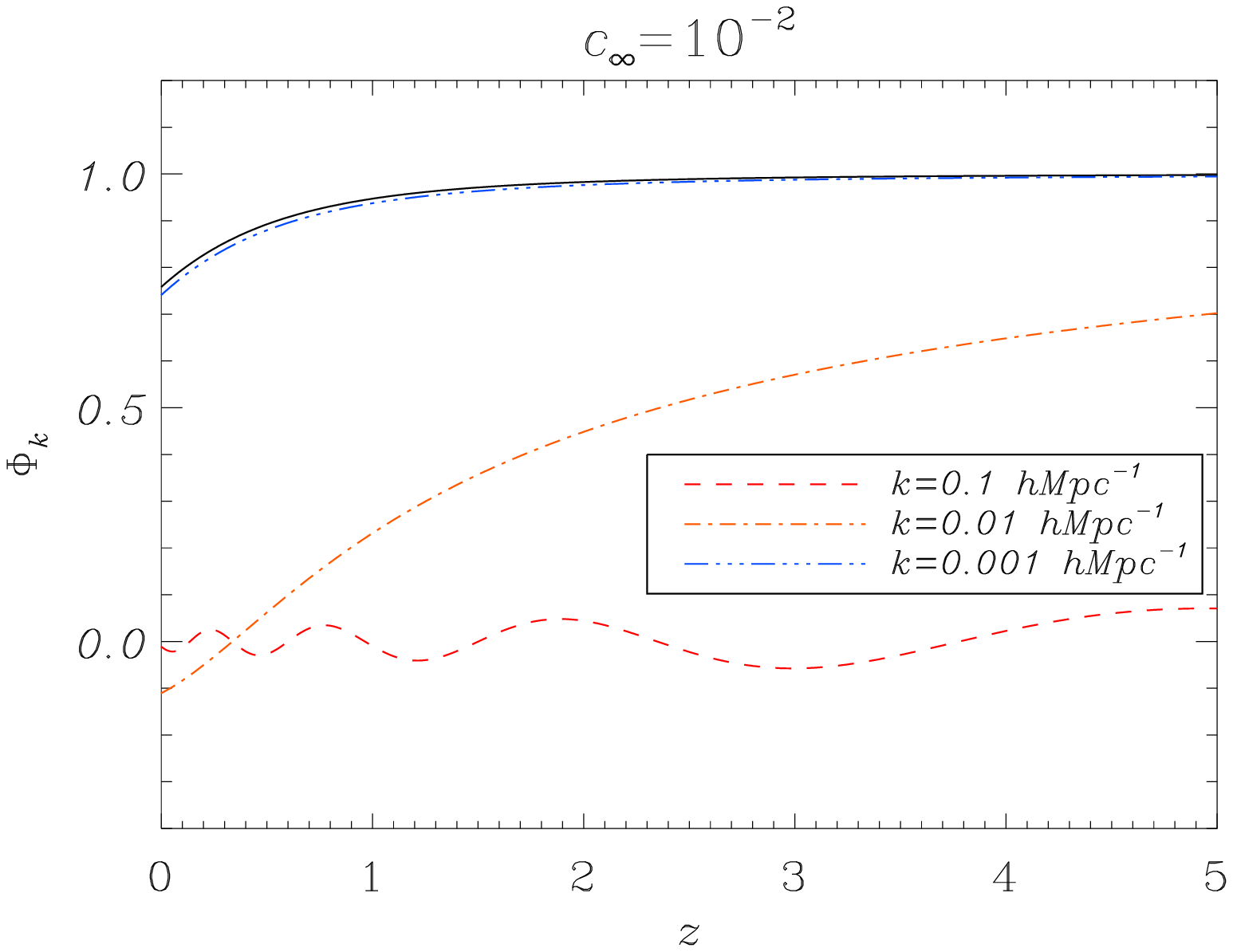}
\includegraphics[width = .45\textwidth]{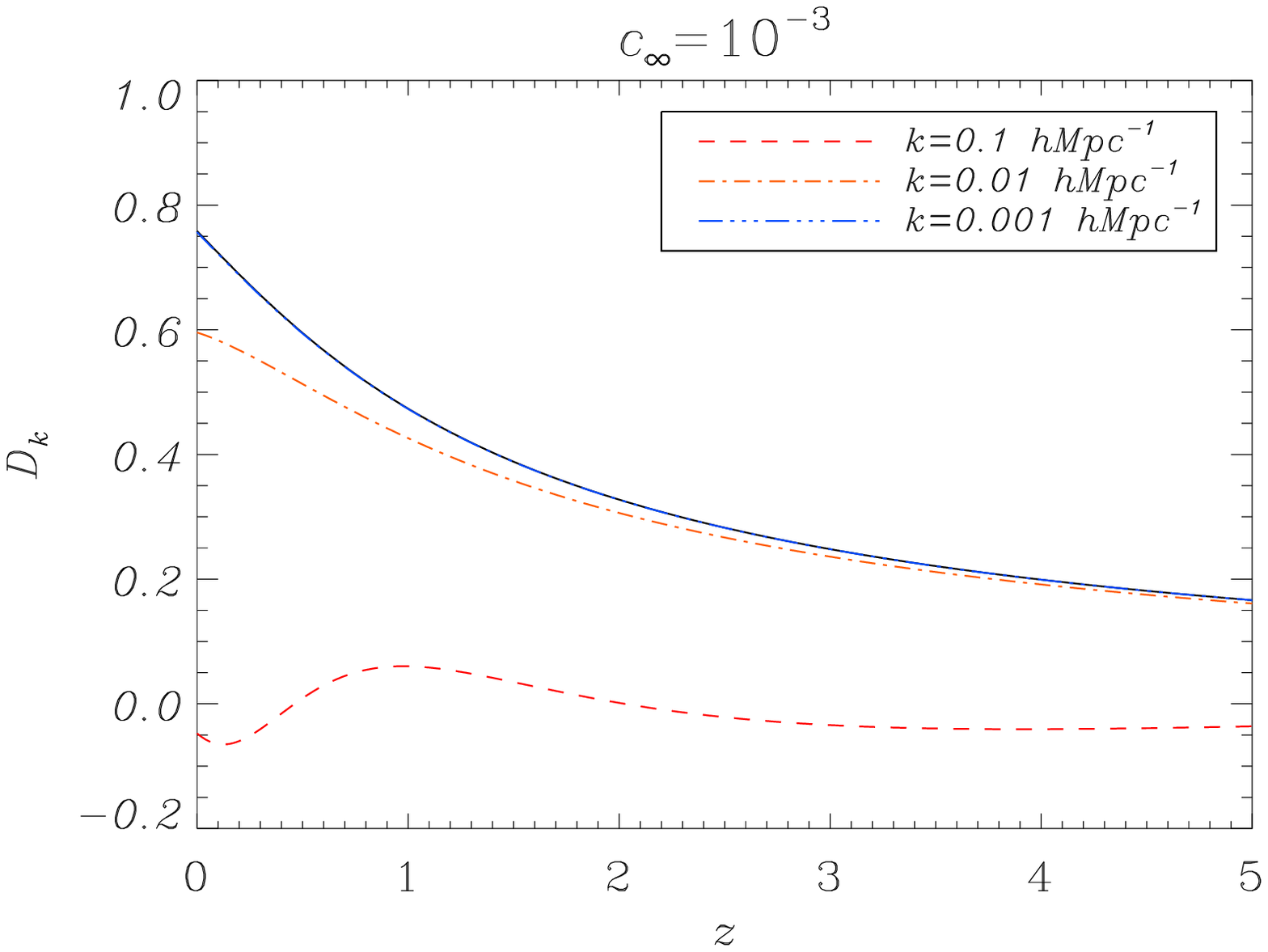}
\includegraphics[width = .45\textwidth]{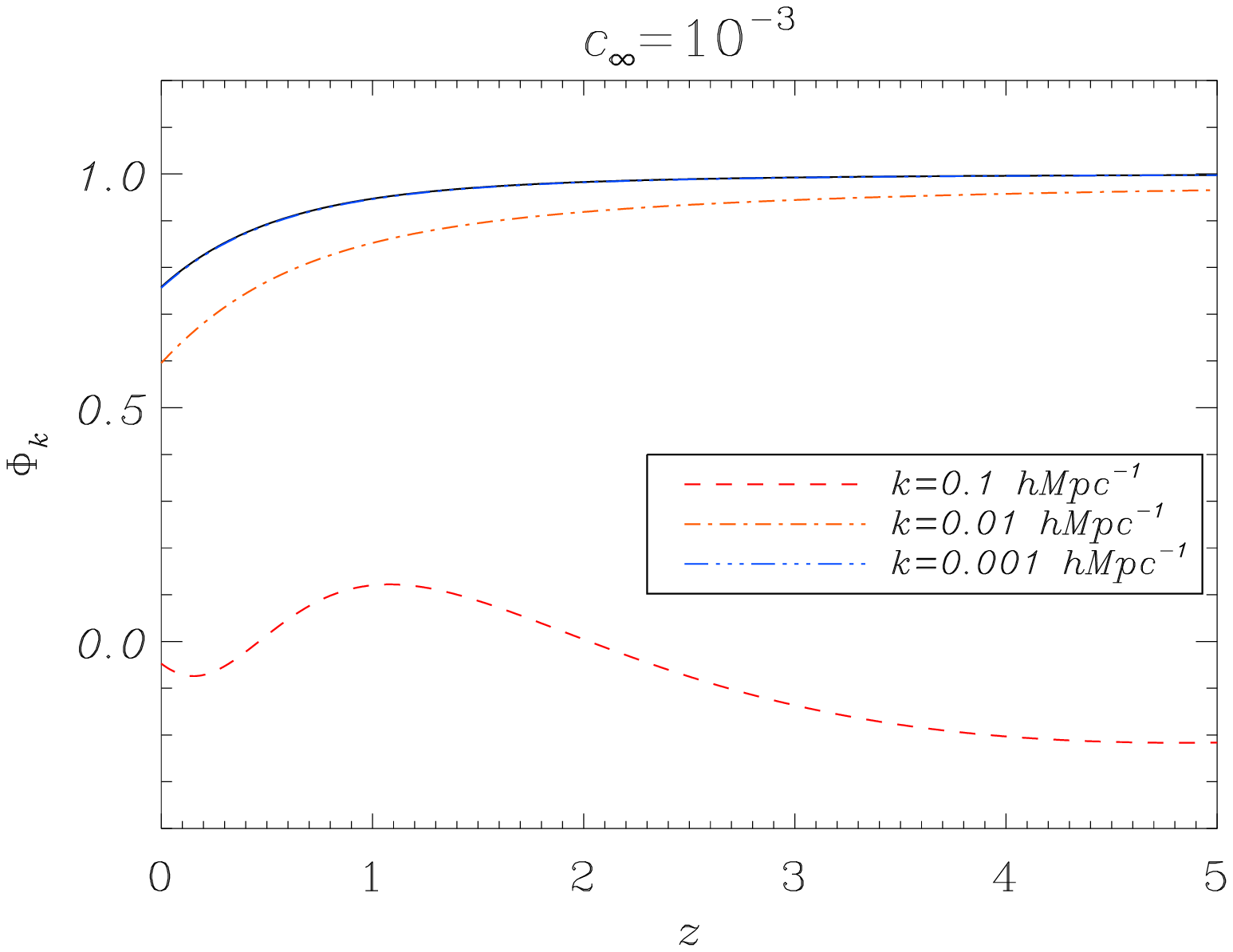}
\includegraphics[width = .45\textwidth]{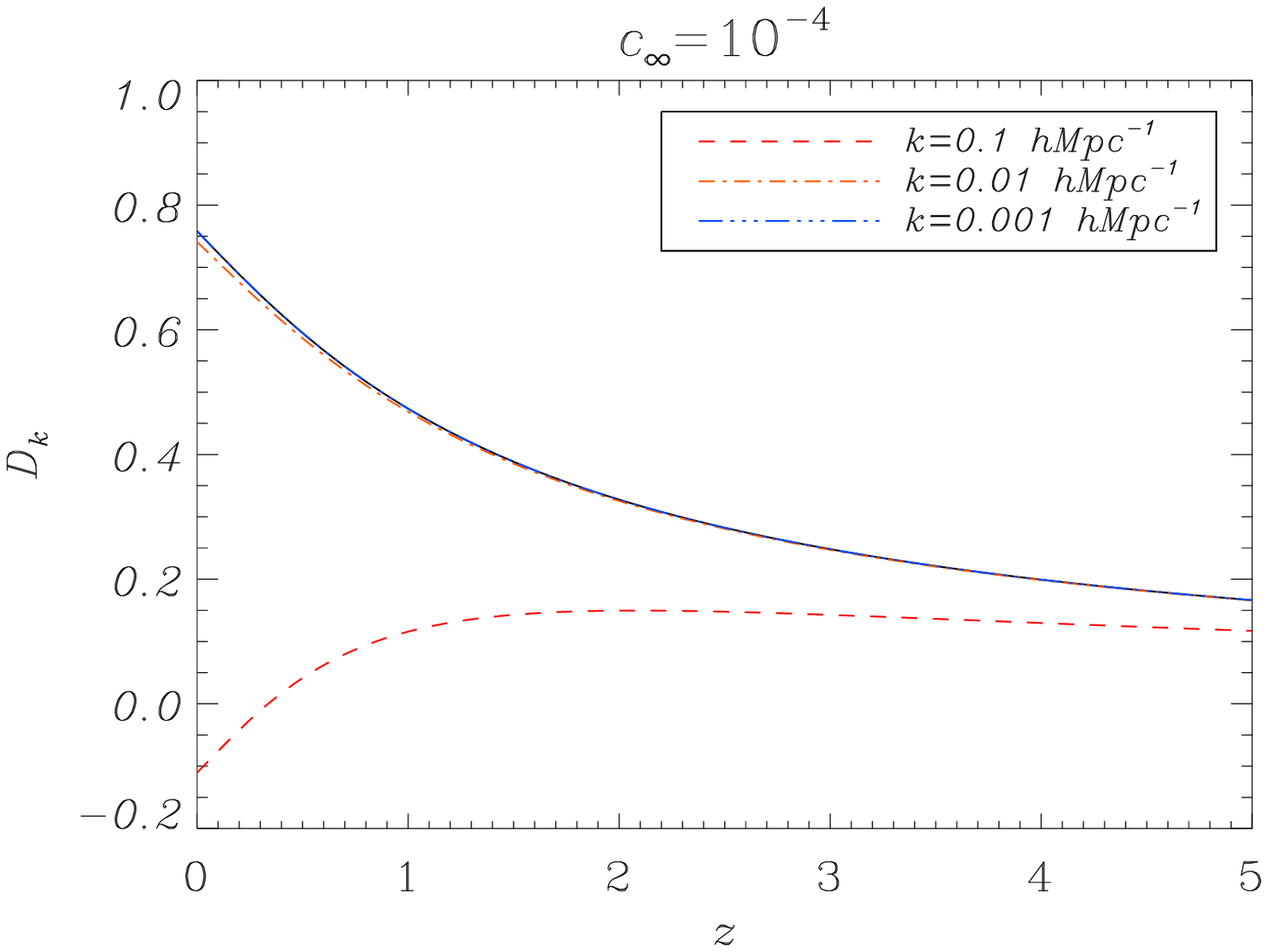}
\includegraphics[width = .45\textwidth]{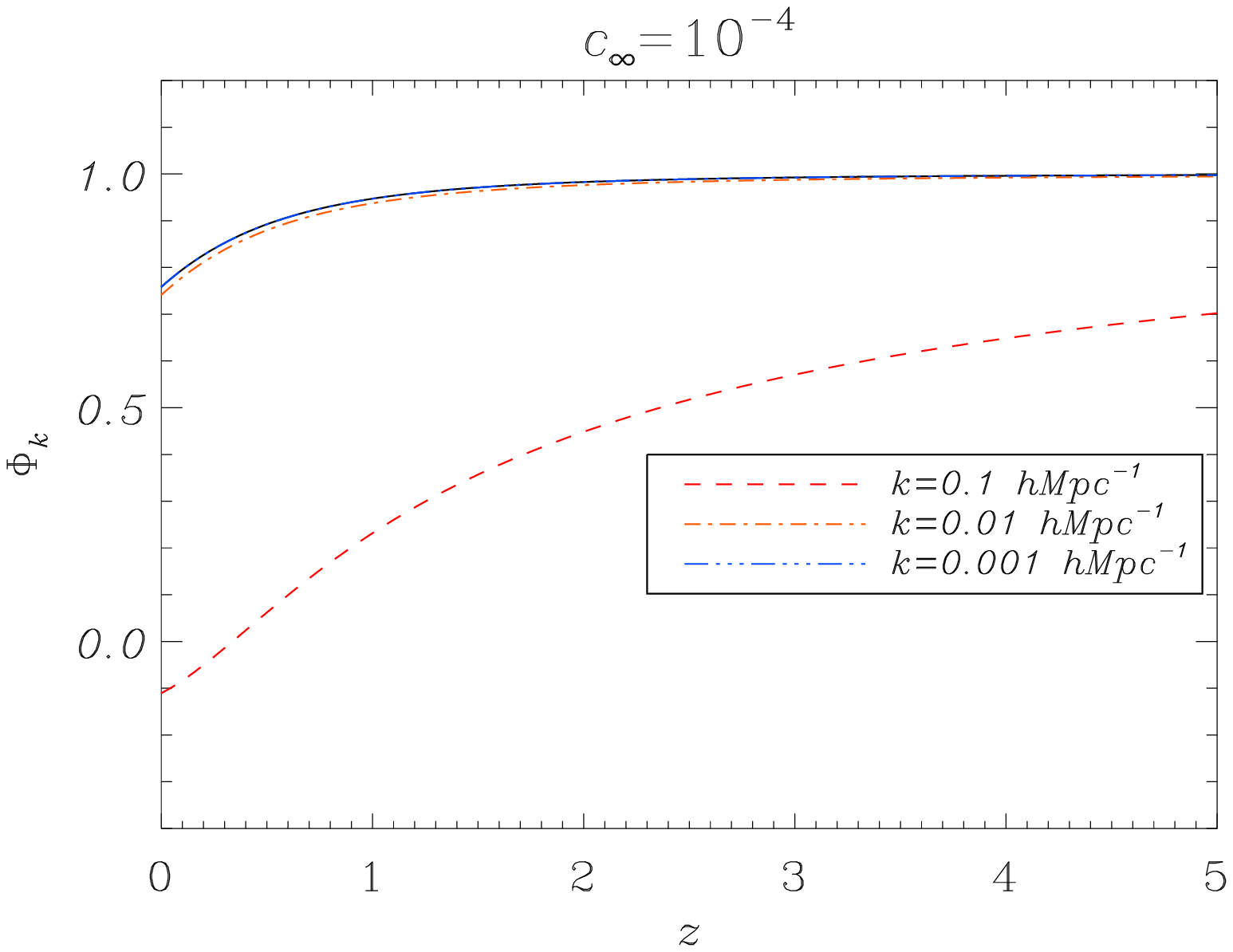}
\caption{Growth factor $D_{\rm k}(z)$ (left column) and normalized gravitational potential  $\Phi_{\rm k}(z)=\Phi({\bf k}; z) /\Phi({\bf 0};10^3) $ (right column) as function of redshift for different Fourier modes. Top to bottom, the sound speed has been set to $c_\infty^2 = 10^{-2}, 10^{-3}, 10^{-4}$. For $k<10^{-3}~{\rm h~Mpc^{-1}}$ the UDM follows the $\Lambda$CDM behaviour (black solid line). For larger modes the discrepancy increases with the sound speed.}
\label{fig:D-phi}
\end{center}
\end{figure}

In Fig.~\ref{fig:A1} we show  the evolution of $T_{\rm UDM}$ as a function of the scale factor for $k = 0.1 \;h$ Mpc$^{-1}$ and of the wavenumber for $a=1$. In order to investigate their behaviour at different scales, we study the evolution of the growth factor and the gravitational potential for some values of $k$. In Fig.~\ref{fig:D-phi} we compare the $\Lambda$CDM case with the UDM one with $c_{\infty}^2 = 10^{-2}, 10^{-3}, 10^{-4}$; for the smallest value of $c_{\infty}^2$, the behaviour coincides with the one of the $\Lambda$CDM model, and the difference from $\Lambda$CDM increases with $k$ (for the gravitational potential plots see also Ref.\ \cite{Camera:2009uz}).
These plots show that for large values of $c_\infty$ (when Jeans length becomes very large), the structure formation is affected and becomes unstable, and this will cause a decrease in the cross-correlation.

\subsubsection{Analytic approximation}
We found a simple analytical expression for the UDM transfer function which fits very well the numerical results given by (see also~\cite{transfer-function}) 
\begin{equation}\label{fittfunc}
T_{\rm UDM}(k;\eta) = j_0( \mathcal{A}(\eta) k) \;,
\end{equation}
where
\begin{equation}\label{Aformula}
\mathcal{A}(\eta) = \int_{\eta_{\rm rec}}^{\eta}c_{\rm s}(\eta')d\eta'= c_{\infty}\int^{a}_{a_{\rm rec}}\frac{da}{Ha^2\sqrt{1 + (1-c_{\infty}^2)\nu a^{-3}}}  \;,
\end{equation}
where  $\nu = \Omega_{\rm m0}/\Omega_{\Lambda 0} \approx 0.387$ and $H^2=H_0^2 \Omega_{\Lambda 0}  (1+\nu a^{-3})$. Since $\mathcal{A}(\eta)$ is not analytical and we are going to consider $c_{\infty}^2 \leq 10^{-2}$ \cite{Bertacca:2007cv}, it is convenient to expand $c_{\rm s}$ in Eq.~(\ref{eq:cs2}) in Taylor series near $c_{\infty} = 0$, i.e.
\begin{equation}
c_{\rm s} = \frac{c_{\infty}}{\sqrt{1 + \nu \left(1 + z\right)^3}} + \frac{1}{2}\frac{\nu\left(1 + 
z\right)^3}{\left[1 + \nu \left(1 + z\right)^3\right]^{3/2}}c_{\infty}^3 + O\left(c_{\infty}^5\right)\;,
\end{equation}
where  $\nu = \Omega_{\rm m0}/\Omega_{\Lambda 0} \approx 0.387$.
The approximation
\begin{equation}\label{cs2eff1storder}
c_{\rm s} \approx \frac{c_{\infty}}{\sqrt{1 + \nu \left(1 + z\right)^3}}\;,
\end{equation}
is therefore very good, being the truncation error of order $c_{\infty}^3$. Plugging Eq.~(\ref{cs2eff1storder}) in Eq.~(\ref{Aformula}), changing the integration variable to the scale factor $a$ and choosing $a = 0$ as lower integration limit we find
\begin{equation}\label{Aformula2}
\mathcal{A}(a) \approx c_{\infty}\int^{a}_{0}\frac{da}{Ha^2\sqrt{1 + \nu a^{-3}}} = 
\frac{c_{\infty}}{H_0\Omega_{\Lambda 0}^{1/2}}\int^{a}_{0}da\frac{a}{a^3 + \nu}\;.
\end{equation}
The integration can be performed analytically and the result is:
\begin{eqnarray}\label{Aformula3}
\frac{H_0\Omega_{\Lambda 0}^{1/2}\mathcal{A}(a)}{c_{\infty}} \approx
\frac{\sqrt{3}\pi}{18\nu^{1/3}} + \frac{1}{6\nu^{1/3}}\ln\frac{a^2 - a\nu^{1/3} + 
\nu^{2/3}}{\left(a + \nu^{1/3}\right)^2} + 
\frac{\sqrt{3}}{3\nu^{1/3}}\arctan\left(\frac{\sqrt{3}}{3}\frac{2a - \nu^{1/3}}{\nu^{1/3}}\right)\;. \nonumber\\
\end{eqnarray}

\begin{figure}[h]
\begin{center}
\includegraphics[width = 0.8 \textwidth]{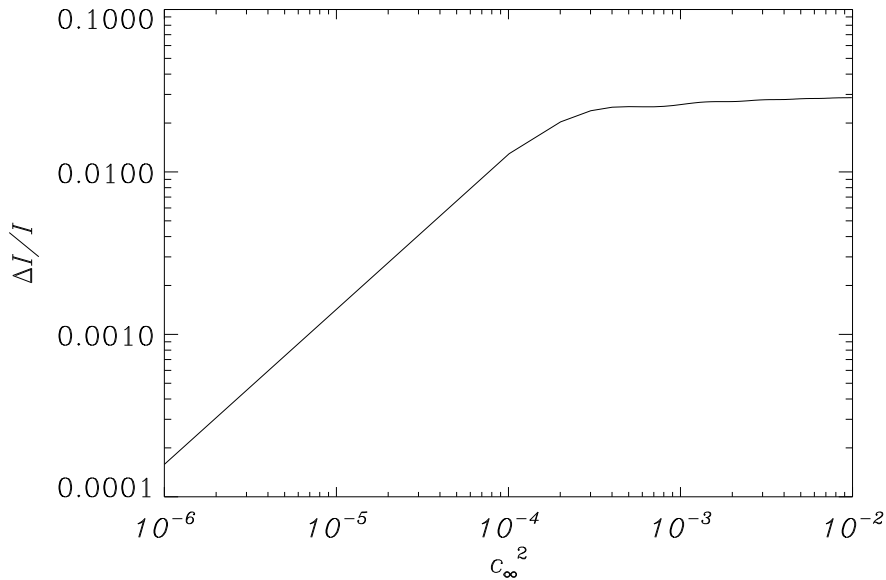}
\caption{Percentage difference between numerical solution and the analytic approximation used in our analysis ($\Delta I/I$, Eq.~(\protect\ref{DeltaI})) as function of $c_\infty^2$. It can be seen that the analytic approximation is very good and the discrepancy with respect to the exact solution is at most few percent.}
\label{fig:DeltaI-rel}
\end{center}
\end{figure}

In order to check the accuracy of  the approximation of our  fitting function \eqref{fittfunc}, let us define the following relation:
\begin{equation}\label{DeltaI}
\frac{\Delta I}{I}(c_\infty^2)=\frac{\left| I_{\rm Fit}-I_{\rm Numeric}\right|}{I_{\rm Numeric}}=\frac{\left|\left[\int T_{\rm UDM}(k;a) dk~da\right]-\left[\int ( \Phi/\Phi_{k \ll 1/\lambda_{\rm J}})(k;a) dk~da\right]\right|}{\left[\int ( \Phi/\Phi_{k\ll 1/\lambda_{\rm J}})(k;a) dk~da\right]}\;.
\end{equation}
From Fig.\ \ref{fig:DeltaI-rel} we can notice that at most we have a relative error about $3\%$ for $c_\infty^2=10^{-2}$.  

Now, using the transfer function of Eq.~(\ref{fittfunc}), 
let us analyse in detail the behaviour of the gravitational potential, the growth factor and the power spectrum of $\delta_{\rm {DM}}$. 

\begin{figure}[h]
\begin{center}
\includegraphics[width = 1 \textwidth]{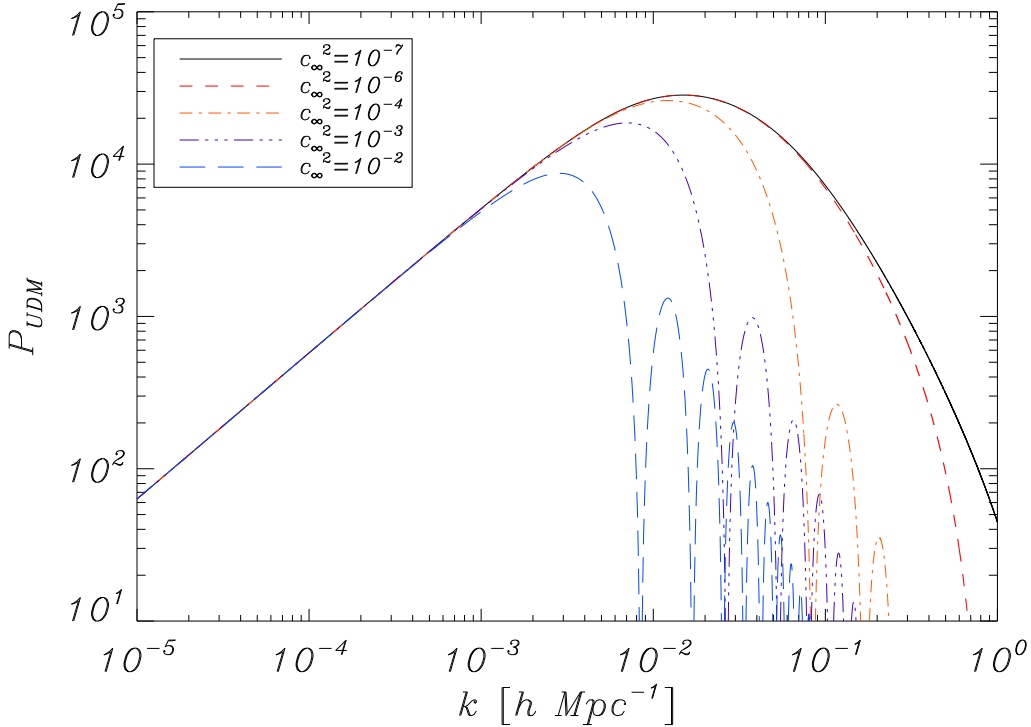}
\caption{Power spectrum $P(k)$ of the UDM models for  a wide range of $c_{\infty}^2$.} 
\label{P_k}
\end{center}
\end{figure}

In Fig.~\ref{P_k} we show the power spectrum of the UDM energy density component that clusters:
\begin{equation}
P(k; z)=\frac{9}{25} T_{\rm m}^2(k) T_{\rm UDM}^2(k;z) \left(\frac{D(z)}{\Omega_{\rm m 0}}\right)^2\left(\frac{k}{H_0}\right)^4P_\Phi(k)\;,
\end{equation}
where 
\begin{equation}
P_\Phi(k)= \left(\frac{50\pi^2}{9}\right) A\delta_H^2 \left(\frac{\Omega_{\rm m 0}}{D(z=0)}\right)^2 k^{-3} \left(\frac{k}{H_0} \right)^{n-1}
\end{equation}
 is the primordial 3D power spectrum of the potential field,
\begin{eqnarray}
\langle \Phi_{\rm p}({\bf k}) \Phi_{\rm p}({\bf k'}) \rangle = (2\pi)^3 \delta({\bf k}+{\bf k}') P_\Phi(k)\;,
\end{eqnarray}
 $\delta_H$ is the expression given by Eq. (A3) of Ref.\ \cite{Eisenstein:1997ij} and $A=1.06$ is computed normalizing the power spectrum to $\sigma_8=0.78$ (see \cite{Raccanelli:2008kx}).

Fig.\ \ref{P_k} confirms that, when $k \gtrsim k_{\rm J}$, $\Phi({\bf k}; z)$, $D(k; z)$ and $P(k; z)$ oscillate and decay both as a function of time and $k$.
As a result, the general trend is that the possible appearance of a sound speed significantly different from zero at late times for these UDM models corresponds to the appearance of a Jeans' length under which the dark fluid does not cluster anymore. At small scales, only if the sound speed is small enough, UDM reproduces $\Lambda$CDM (see also \cite{Bertacca:2007cv, Camera:2009uz,Camera:2010wm}).

\section{Cross correlation between the CMB and the Large Scale Structure} 
\label{3}
The integrated Sachs-Wolfe effect is an energy shift which a photon exhibits when it crosses a gravitational potential evolving in time. Since in a Universe made of pressure-less matter the gravitational potential is constant, in the standard cosmological scenario we may distinguish two effects: one at early time, \emph{early} ISW effect, during the transition from a radiation-dominated Universe to a matter dominated one, and one at late time, \emph{late} ISW effect when the evolution of the Universe starts feeling the presence of a cosmological constant. 
The late ISW takes place at recent epoch and affects large cosmological scales: this reflects into a bump at low multipoles in the angular power spectrum of the CMB. It is well known that this region has a large intrinsic uncertainty due to the cosmic variance and the constraining power results pretty low. The cosmological constant affects the structure formation at large scales as well: cross-correlating the distribution of galaxies with the CMB \cite{Crittenden:1995ak} has been proven to increase the signal-to-noise ratio \cite{Boughn:2001zs, Boughn:2003yz} and to be a useful probe of the late time evolution of the Universe. 

In particular this is true in the case of UDM models. Indeed
it is important to stress that in the UDM models there are two
simple but important aspects: first, the fluid which triggers the
accelerated expansion at late times is also the one which has to
cluster in order to produce the structures we see today. Second,
from the last scattering
to the present epoch, the energy density of the Universe is
dominated by a single dark fluid, and therefore the gravitational
potential evolution is determined by the background and
perturbation evolution of just such a fluid. As a result, the possible appearance of a sound speed
significantly different from zero at late times corresponds to the
appearance of a Jeans length (or a sound horizon) under which the
dark fluid does not cluster anymore, causing a strong evolution
in time of the gravitational potential and of the fractional overdensity $\delta_{\rm DM}$ which start to oscillate
and decay. Thus, besides having the possibility of a strong ISW effect for UDM models \cite{Bertacca:2007cv}, 
it is interesting to analyse the possible changes of the cross-correlation between the distribution of galaxies and CMB  signal in UDM models due to a non-negligible sound speed, for example, 
with respect to the $\Lambda$CDM model.

 Following Refs.~\cite{Hu:2004yd} and \cite{Corasaniti:2005pq}, consider a field $x(\hat{\mathbf{\gamma}})$ as a function of the angular position $\hat{\mathbf{\gamma}}$ on the sky to be a weighted projection of the potential field $\Phi({\bf x};z)$:  
\begin{equation}
\delta x(\hat{\mathbf{\gamma}}) = \int dz ~ W^x(\hat{\mathbf{\gamma}};z) \Phi(\hat{\mathbf{\gamma}};z)\,,
\end{equation}
where the weight $W^x$ can also, in general, include differential operators acting on the field. The angular cross-correlation between two observed fields $x$ and $x'$ is given by
\begin{equation}
c^{xx'} (\vartheta)=\langle \delta x(\hat{\mathbf{\gamma}}) \delta x'(\hat{\mathbf{\gamma}}') \rangle = \sum_\ell  P_\ell(\hat{\mathbf{\gamma}}\cdot\hat{\mathbf{\gamma}}')  {{2\ell+1} \over 4\pi}C_\ell^{xx'}\,.
\end{equation}
The brackets $\langle \rangle$ denote averaging over all direction $\hat{\mathbf{\gamma}}$ and $\hat{\mathbf{\gamma}}'$, satisfying the
condition $\hat{\mathbf{\gamma}}\cdot \hat{\mathbf{\gamma}}'=\cos(\vartheta)$, 
$P_\ell$ is the Legendre polynomial  and $C_\ell^{xx'}$ is the angular cross power spectrum, which is given by
\begin{equation}
C_\ell^{xx'} = 4\pi \int {d^3 k \over (2\pi)^3} ~ I_\ell^x(k) I_\ell^{x'}(k) P^{\Phi\Phi}\,,
\end{equation}
where the weight $I_\ell^x(k)$ is given by
\begin{equation}
I_\ell^{x}(k) =\frac{9}{10}T_{\rm m}(k) \int dz ~ (1+z)T_{\rm UDM}(k; z) D(z) W^x(k;z) j_\ell (k \chi(z))\;,
\end{equation}

and $j_\ell (k \chi(z))$ are the spherical Bessel functions. Here $\chi(z) = \int_0^z dz/H(a)$ is the comoving distance to redshift $z$.  

When cross-correlating CMB and LSS to measure the ISW effect, $\delta x = \delta T=\Theta$, where $\Theta$ is  the CMB temperature fluctuation,  and $\delta x^\prime = \delta_{n_g}(\mathbf{\hat\gamma},z)$,  where $\delta_{n_g}=\delta n_g/n_g$ is the galaxy overdensity observed and where $n_g(\mathbf{\hat\gamma},z)$ is the number of sources in the sky which 
traces the matter distribution \cite{Nolta:2003uy}. 
The window functions read
\begin{eqnarray}
&& W^{T}(k ;z) =- 2 T_{\rm CMB} ~ e^{-\tau(z)}{\partial \ln \left[ (1+z)T_{\rm UDM}(k; z) D(z) \right] \over \partial z}\; \\
&& W^{n_g}(k;z)=-\frac{ b(z) k^{2}}{\left(3/2\right)H_0^2\Omega_{\rm m0}\left(1 + z\right)} \frac{d N(z)}{d z}\;,
\end{eqnarray}
where $e^{-\tau(z)}$ is the visibility function, which accounts for the effect of reionization; 
$b(z)$ is the galaxy bias and $dN(z)/dz$ is the mean number of galaxies per steradian with redshift $z$ within $dz$. Therefore, the integrand functions $I^{T}_l(k)$ and $I^{n_g}_l(k)$ are respectively:
\begin{eqnarray}
I^{T}_l(k)&=&-\frac{9}{5}T_{\rm CMB} ~ T_{\rm m}(k) \int dz ~ e^{-\tau(z)}{\partial \left[ (1+z)T_{\rm UDM}(k; z) D(z) \right] \over \partial z} j_l[k \chi(z)] \;, \\
I^{n_g}_l(k)&=&-\frac{3}{5} \frac{T_{\rm m}(k) k^{2}}{H_0^2\Omega_{\rm m0}} \int dz ~ b(z) \frac{dN(z)}{d z} T_{\rm UDM}(k; z) D(z) j_l[k \chi(z)] \label{growth}\;.
\end{eqnarray}

\begin{figure}[h]
\begin{center}
\includegraphics[width = .45\textwidth]{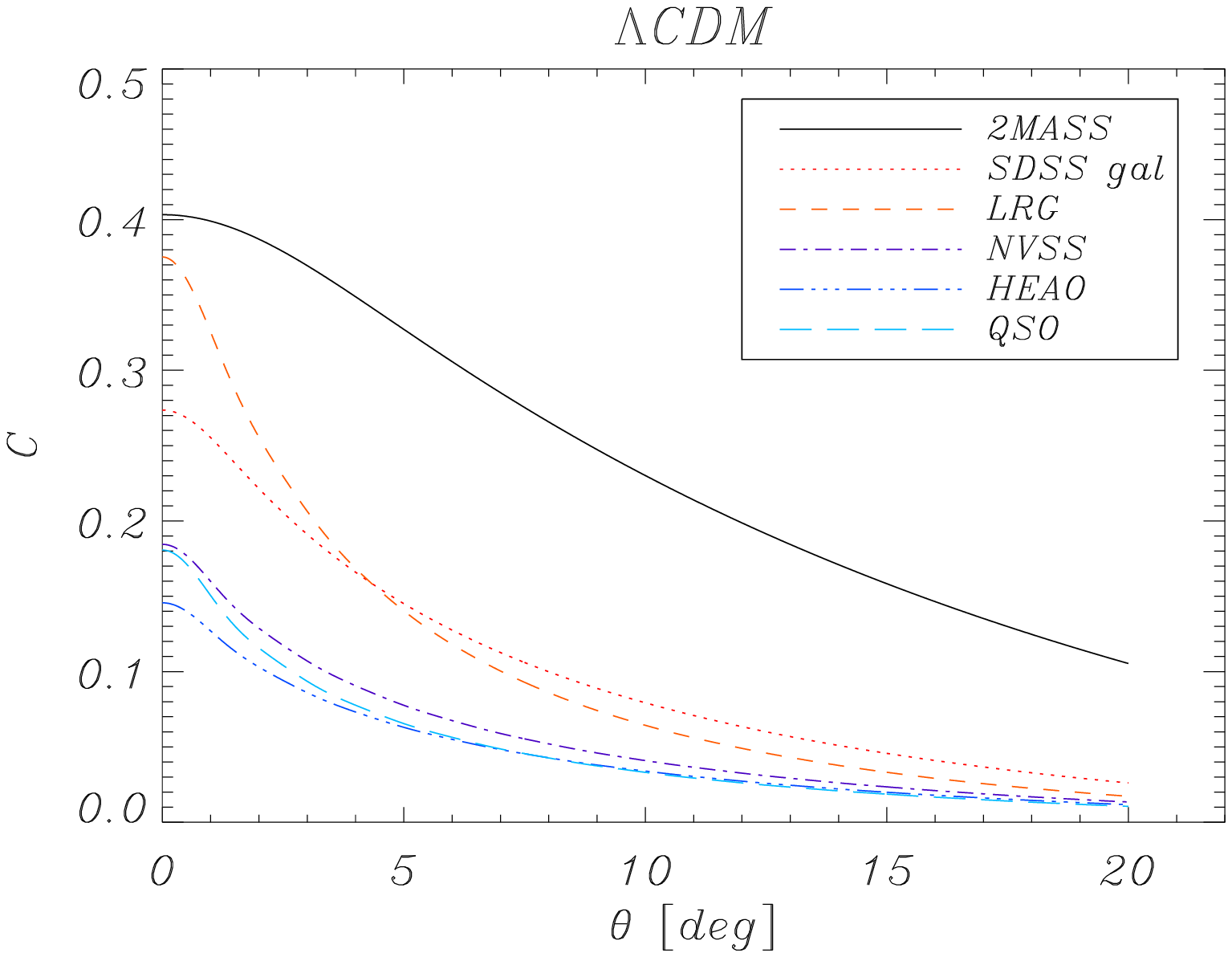}
\includegraphics[width = .45\textwidth]{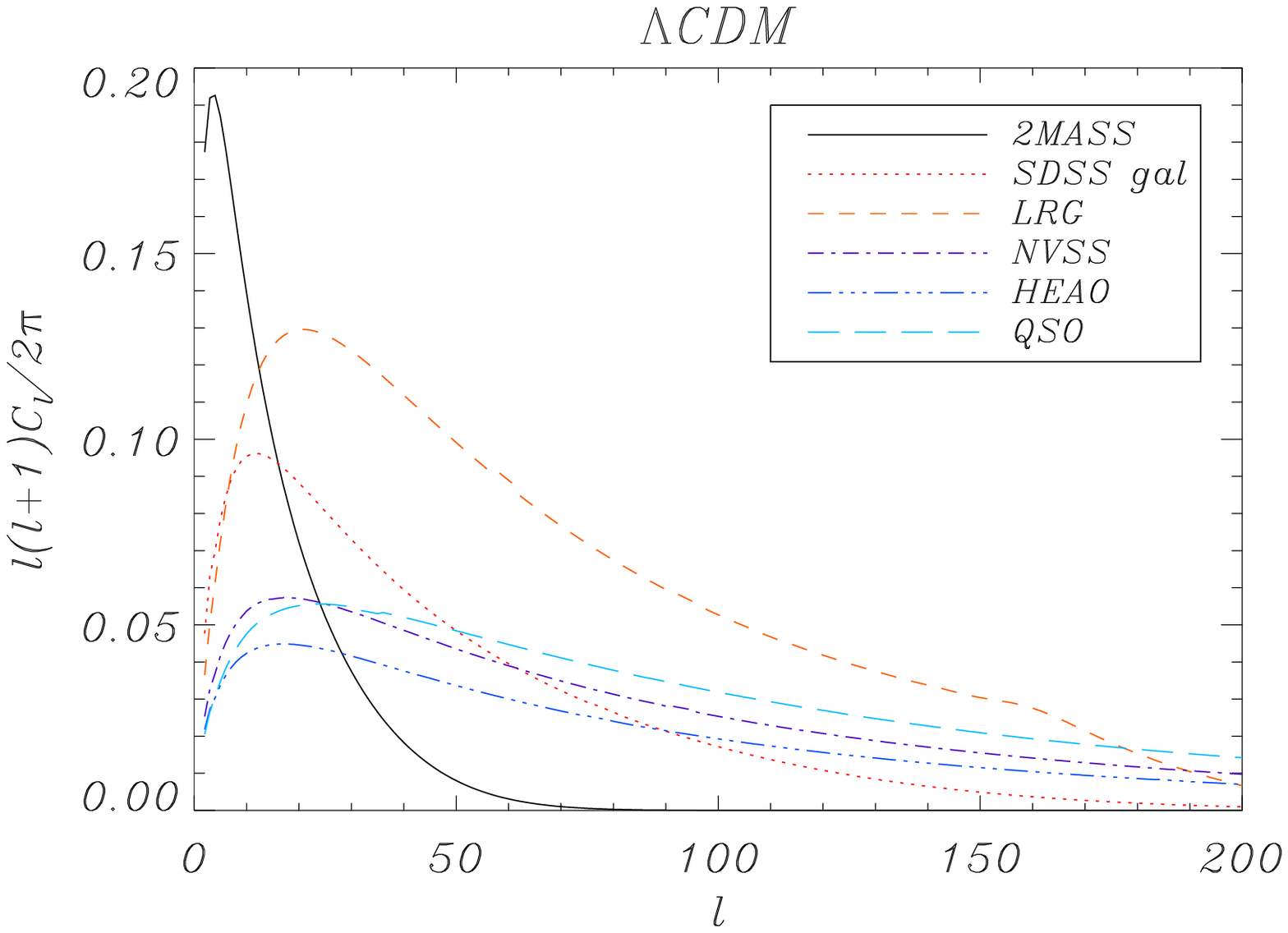}
\includegraphics[width = .45\textwidth]{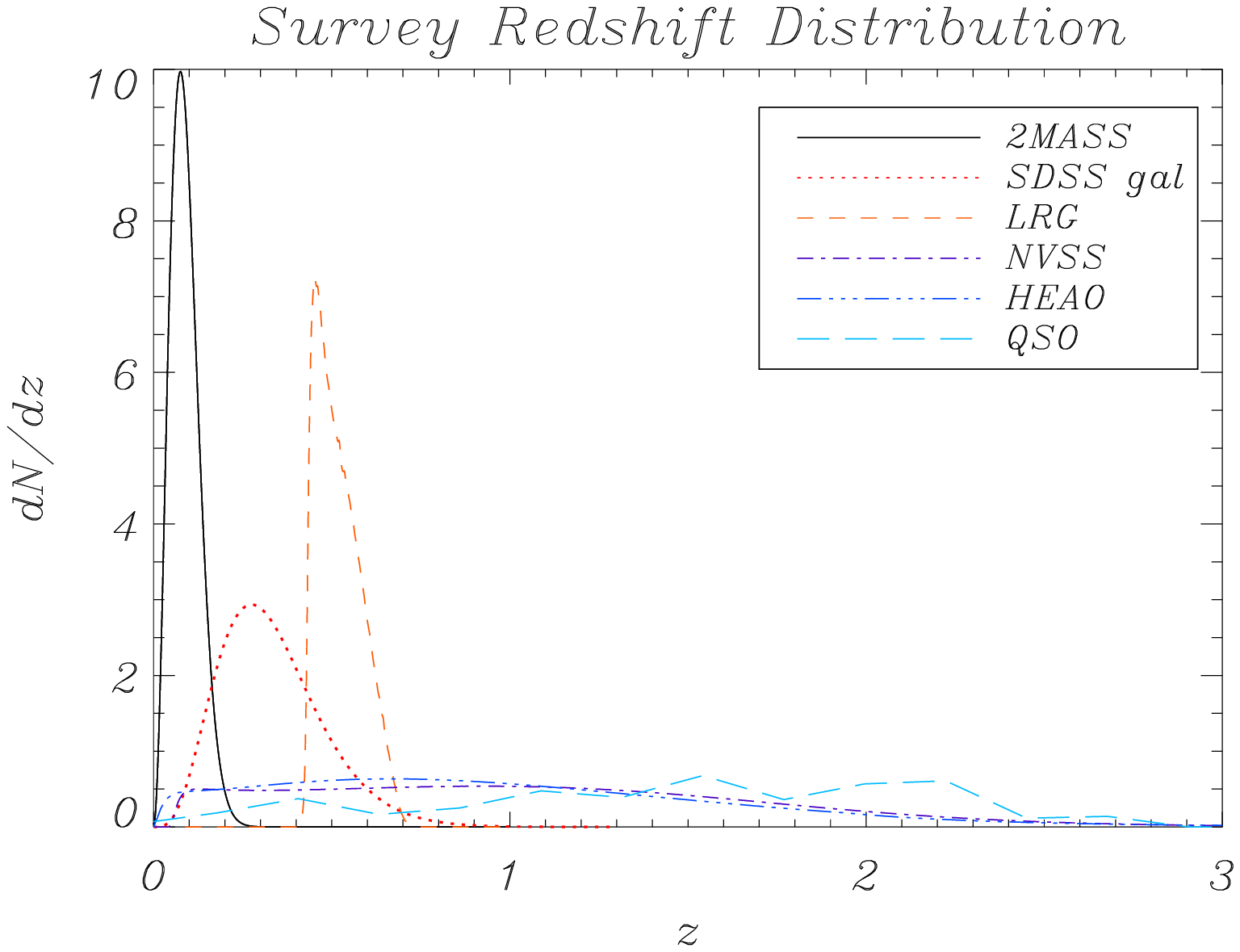}
\includegraphics[width = .45\textwidth]{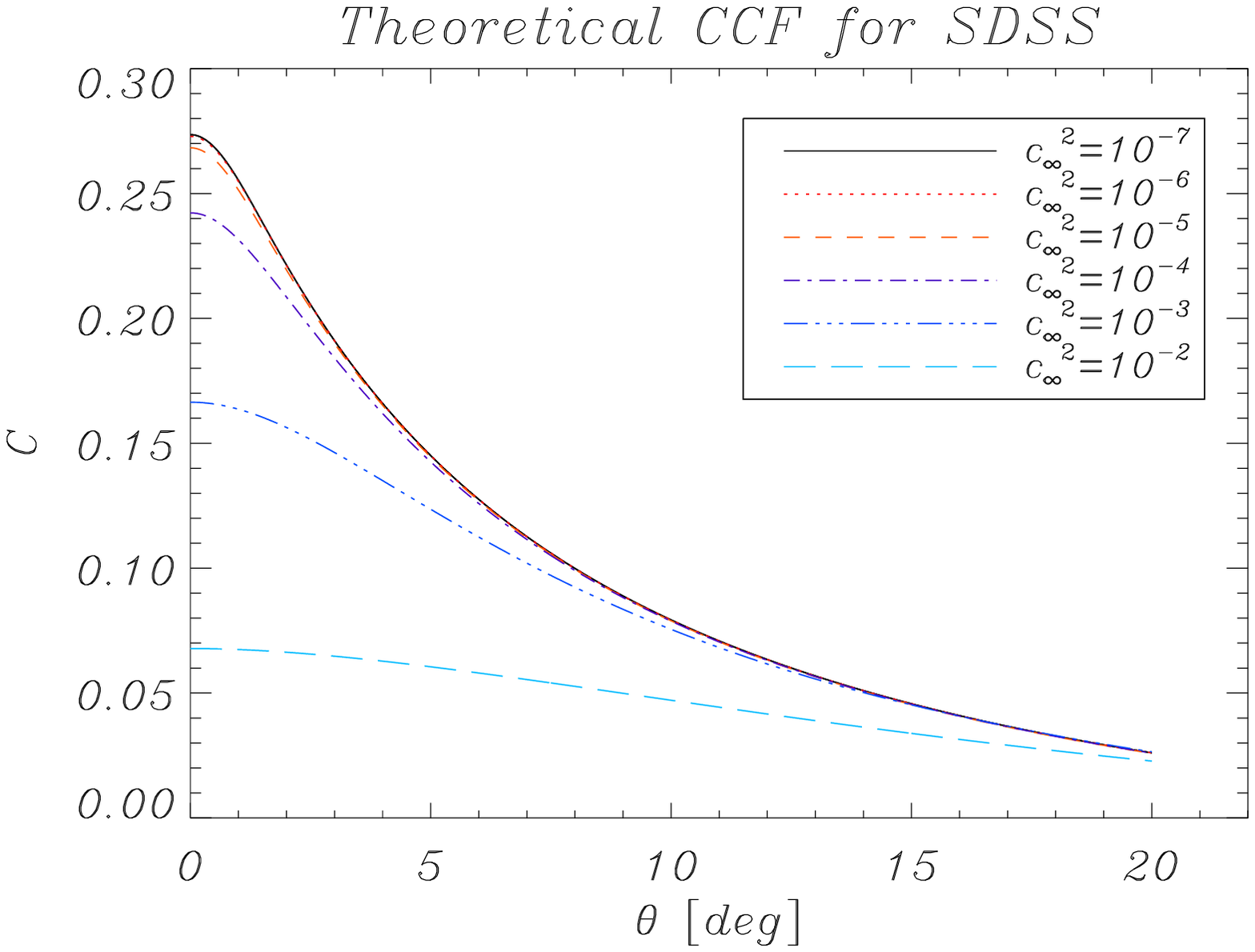}
\caption{\emph{Top}: cross-correlation function between WMAP temperature CMB map and 6 catalogues of sources both in real space (\emph{left panel}) and harmonic space (\emph{right panel}). The redshift distribution of each catalogue is shown in the bottom-left panel whereas the effect of $c_{\infty}^2$ is shown for the specific case of the SDSS galaxy survey (bottom-right).}
  \label{fig:ccf}
\end{center}
\end{figure}

Finally, the Legendre transform of the cross power spectrum $C^{Tn_g}_l$ is the angular cross-correlation function, which can also be defined as the average
\be
\label{cTn_g}
c^{Tn_g} (\vartheta)  = \langle \Theta(\mathbf{\hat  \gamma}_1) \, \delta_{n_g}(\mathbf{\hat \gamma}_2)   \rangle  = b \, \langle \Theta(\mathbf{\hat \gamma}_1) \, \delta_m(\mathbf{\hat \gamma}_2)   \rangle,
\ee
where the CMB temperature fluctuation $\Theta$ and the galaxy overdensity $\delta_{n_g}$ are observed in any directions $\mathbf{\hat \gamma}_1, \, \mathbf{\hat \gamma}_2 $ separated by an angle $\vartheta$. For simplicity, we have assumed that the galaxy bias  $b$ is constant for each data set. The theoretical cross-correlation functions for the large scale surveys discussed in \cite{Giannantonio:2008} (NVSS, SDSS main galaxies, LRGs, and quasars, HEAO, 2MASS) are plotted in Fig.~\ref{fig:ccf}. The actual shape of each catalogue depends on the redshift distribution of the sources,  as shown in top panels both for the two-point correlation function and its Legendre transformation, however in all cases considered, the UDM model with $c_\infty^2 \leq 10^{-4}$ is extremely close to the $\Lambda$CDM one. In the right lower panel, the particular case of SDSS main galaxies catalogue is computed for different values of the parameter $c_\infty^2 = 10^{-2},\dots,10^{-7}$ to highlight the effect of the speed of sound on the CCF. The bottom left panel displays the redshift distributions of the catalogues to help understanding their properties.

The ISW effect is sensitive to the time derivative of the gravitational potential, hence for the UDM model we discuss, it is expected to be higher for large values of the sound speed: the cross-correlation signal in these cases is lower than in the $\Lambda $CDM model. This effect can be explained through the oscillatory behavior of the density contrast and of the gravitational potential when $k \gtrsim k_{\rm J}$ (see Eqs.~(\ref{Jeans}), (\ref{u-cs_k>>theta''/theta}) and (\ref{delta-dm})).
Indeed this is due to two possible factors: {\it i)} through the amplitude of the density contrast that decays over time: the presence of a non-negligible speed of sound prevents the structure formation on scale smaller than the Jeans' length (see Eq.~\ref{Jeans}), so that one expects a cross-correlation in these cases lower than in the $\Lambda $CDM model;
{\it ii)} the rapid fluctuations between positive and negative values produce a temporal mismatch between $\Phi'$ and $\delta_{\rm DM}$ which determines a lowering of the cross-correlation signal. 

In the next section, we perform a $\chi^2$ analysis to determine which value of the sound speed fits better the data. The measured 2-point correlation functions as well as the covariance matrix between the catalogues are publicly available\footnote{http://www.usm.uni-muenchen.de/\~~tommaso/iswdata.php}, whereas the theoretical curves are computed implementing the algorithm described in \cite{Raccanelli:2008kx} following the specifics in \cite{Giannantonio:2008}. This analysis is not meant to fully span the parameter space of the model, which is beyond the purpose of this paper. We assume the WMAP best-fit flat cosmology \cite{Dunkley:2009} for $\Omega_{\rm b}h^2$, $\Omega_{\rm c}h^2$, ${\rm H_0}$, $n_{\rm s}$, $A_{\rm s}$ and $\Omega_\Lambda$ and vary $c_\infty$ only, which is a peculiar feature of the UDM model under investigation.

\section{Data analysis}
\label{sec:data_analysis}

We use the ISW data from \cite{Giannantonio:2008}, which were obtained cross-correlating six galaxy catalogues with the WMAP maps of the CMB. This data set probes the distribution of the large-scale structure in a wide redshift range $ 0 < z < 2 $, using different parts of the electro-magnetic spectrum. The used catalogues are: the 2MASS infrared survey, the optical SDSS, which includes main galaxies, luminous red galaxies and quasars, the radio-galaxies from NVSS and the X-ray background of HEAO. Although not all the redshift distributions are known with a great accuracy, this represents a first approximation towards a true tomographic reconstruction of the potentials.

All maps were coarsely pixellated on the sphere with a resolution of 0.9 deg. Then the real-space cross-correlation functions (CCFs) were measured in angular bins of 1 deg for $0 \, \mathrm{deg} \le \vartheta \le 12 \, \mathrm{deg}$, so that the full data set for the six catalogues consists of 78 points $\left(c_i^{Tn_g} \right)_{\mathrm{obs}}$. 

The full, strongly non-diagonal, covariance matrix $\mathbf{C_{ij}}$ representing the uncertainties on these points was calcolated in three ways, using a model-independent jack-knife (JK) approach and two different Monte Carlo methods. In the first (MC1), the real LSS maps were cross-correlated with 5000 random CMB maps generated using Gaussian random seeds. In the second (MC2), also the LSS maps were randomly produced, based on their redshift distributions. These methods give comparable results, although it was found that the JK approach is rather unreliable, due to strong dependencies on the particular followed procedure. Here we decided to use the MC1, in order to conservatively avoid having to use the redshift distributions of the catalogues in the calculation.

We then calculate the likelihood of a given model by calculating its theoretical CCF $\left(c_i^{Tg} \right)_{\mathrm{theo}}$, based on the measured redshift distributions of each catalogue and on linear theory. The likelihood is then obtained as $\mathcal {L} \propto e^{- \chi^2 / 2}$, where
\be
\chi^2 = \sum_{ij} \left[ \left(c_i^{Tn_g} \right)_{\mathrm{obs}} - \left(c_i^{Tn_g}\right)_{\mathrm{theo}} \right] \, \left[ \mathbf{C^{-1}} \right]_{ij} \, \left[ \left(c_j^{Tn_g} \right)_{\mathrm{obs}} - \left(c_j^{Tn_g}\right)_{\mathrm{theo}} \right] \, .
\ee
We assume that the galaxy bias is constant for each catalogue and we allow it to vary around the value $b_{\mathrm{fid}}$ reported by previous authors and confirmed by \cite{Giannantonio:2008}, in the range $0.5 \, b_{\mathrm{fid}} \le b \le 2 \, b_{\mathrm{fid}}$, and marginalising over this additional degree of freedom. We discuss our findings in the next section.

\section{Results and discussion} 
\label{results}

\begin{figure}[h]
\begin{center}
\includegraphics[width=.8\textwidth]{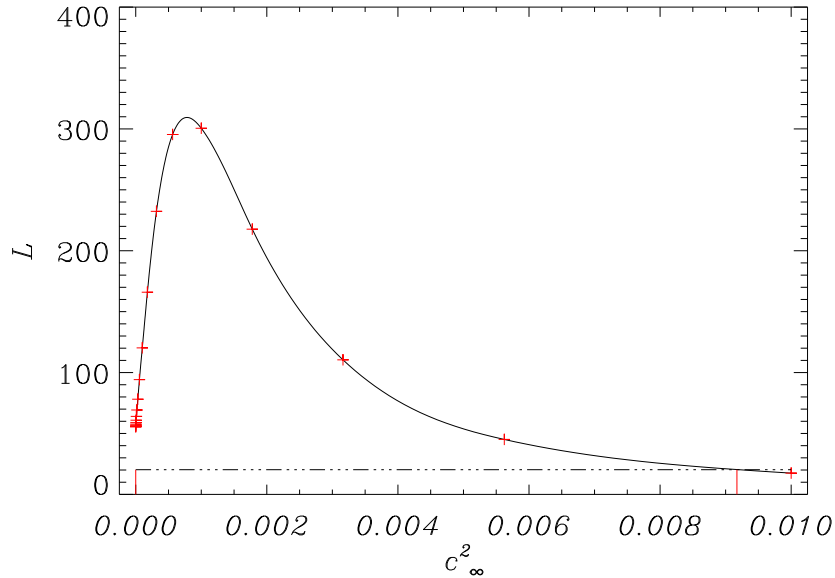}
\caption{Total likelihood of the parameter $c_\infty^2$, in the UDM model resulting from the combined analysis of the publicly available galaxy surveys. The horizontal dot-dashed line marks the 95\% confidence interval.}
\label{fig:cs_like}
\end{center}
\end{figure}

The total likelihood for the speed of sound $c^2_\infty$ is shown in Fig.~\ref{fig:cs_like}. The 95\% confidence level limits are $c^2_\infty\in[0., 0.009]$. This reveals that despite the large number of surveys combined together the constraining power is weak and we hit the priors. The $\Lambda$CDM model is a good fit to the data, but the distribution peaks to the value $c^2_\infty=10^{-4}$. We check the stability of the solution against the bias correction repeating the analysis assuming the fiducial value for each catalogue. While the different treatment of the uncertainty on how baryon trace cold dark matter marginally affects the confidence region at 68\%, the 95\% bounds are unchanged. 

\begin{figure}[h]
\begin{center}
\includegraphics[width=.45\textwidth]{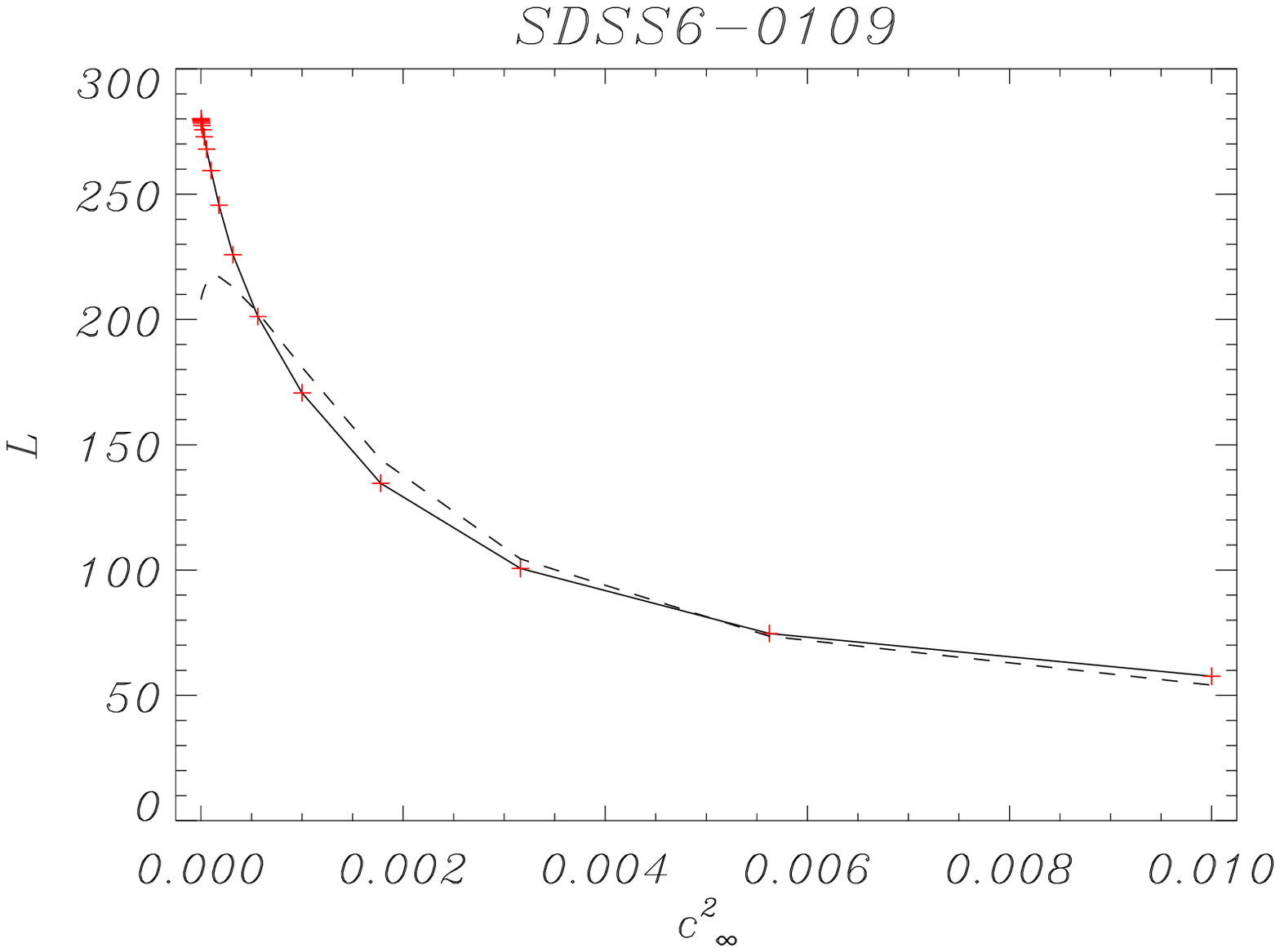}
\includegraphics[width=.45\textwidth]{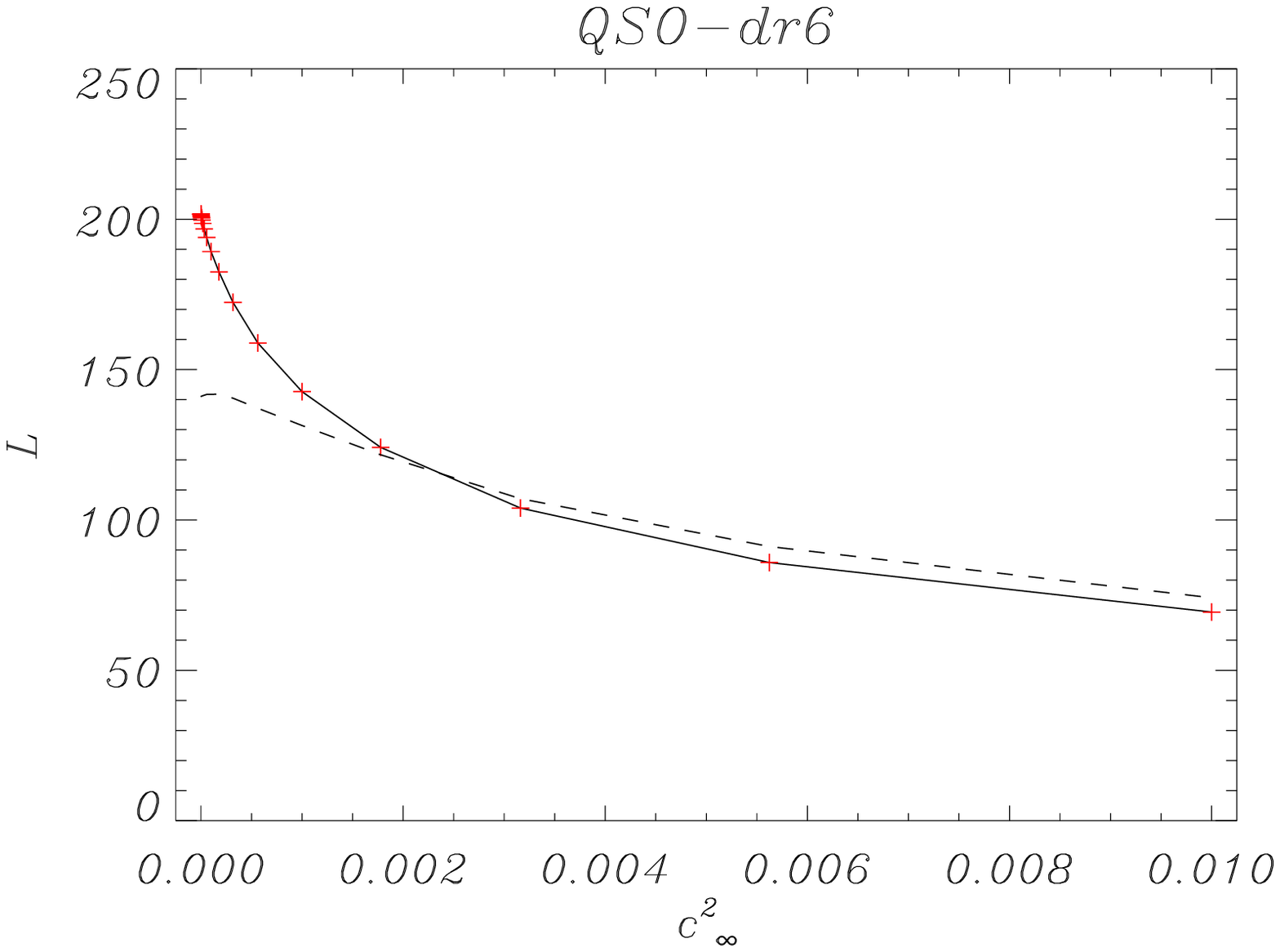}
\includegraphics[width=.45\textwidth]{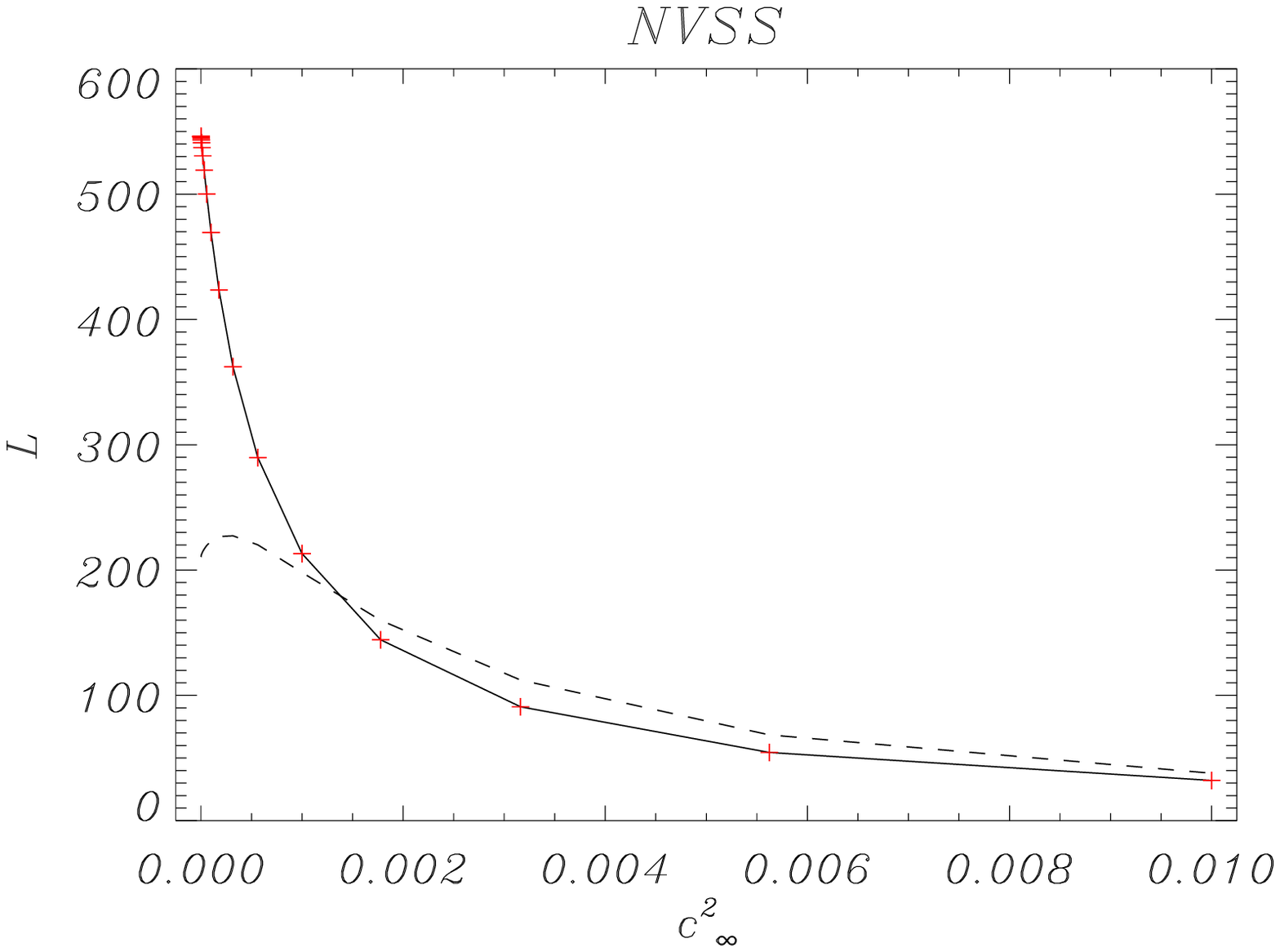}
\includegraphics[width=.45\textwidth]{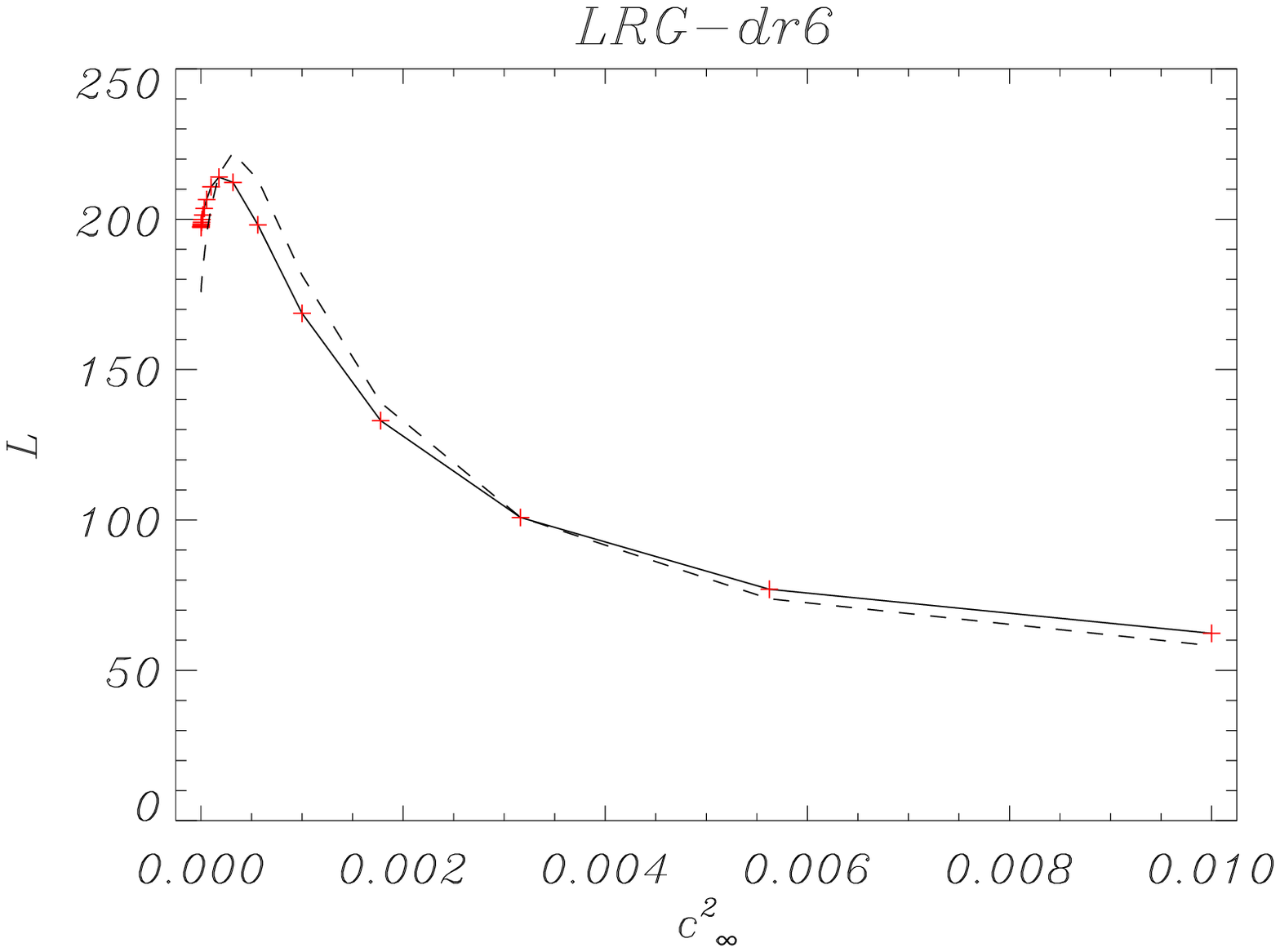}
\includegraphics[width=.45\textwidth]{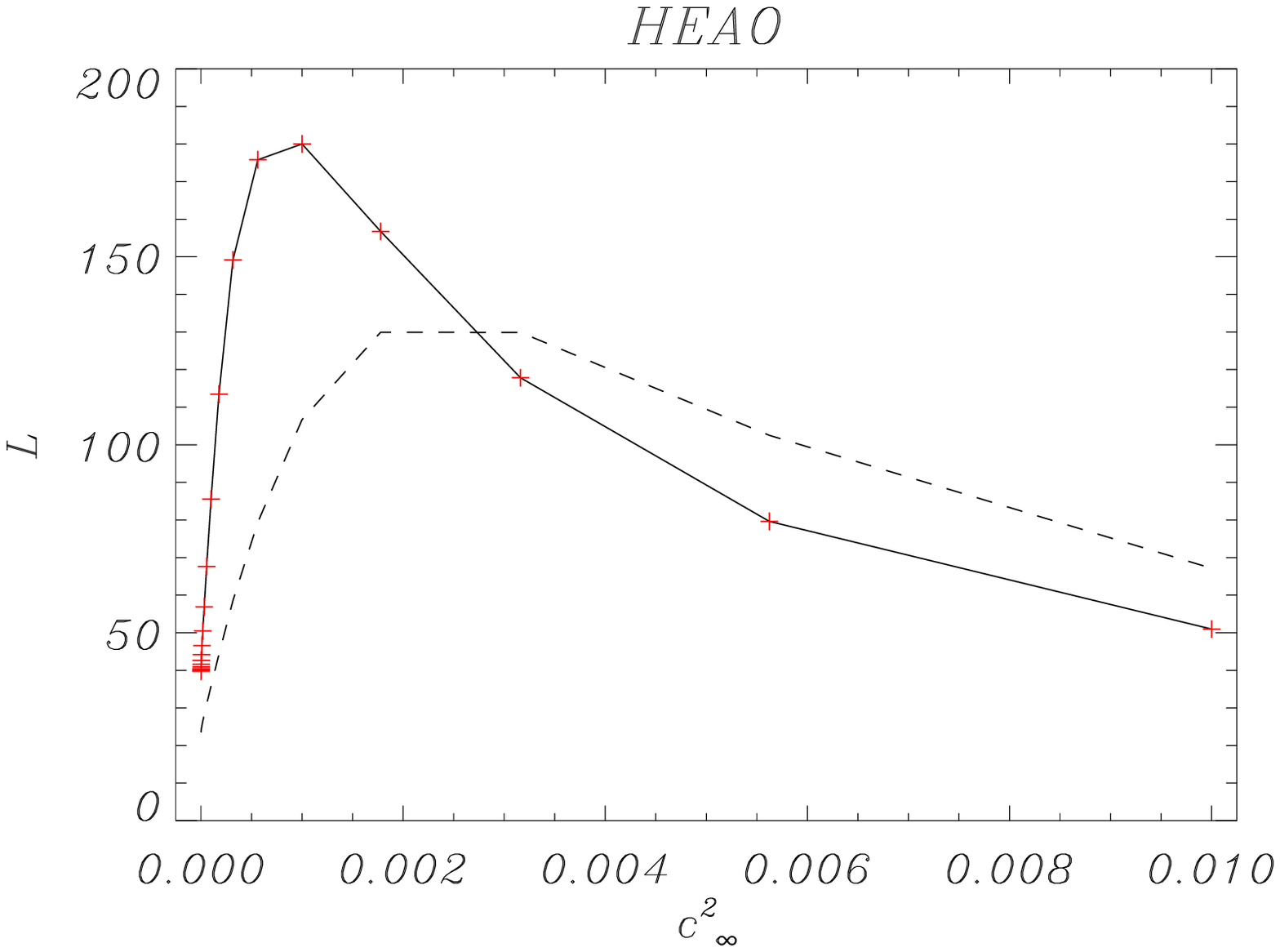}
\includegraphics[width=.45\textwidth]{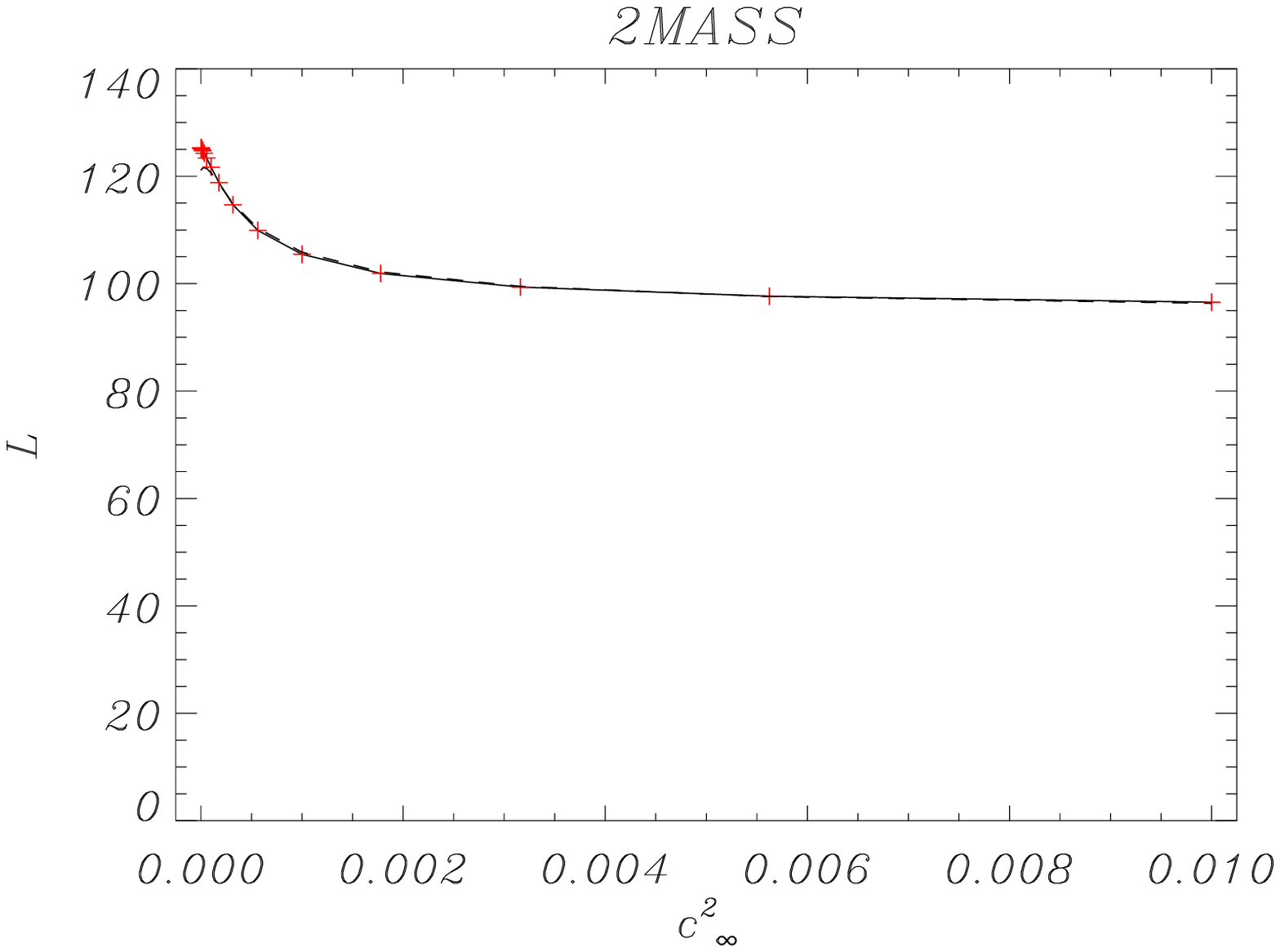}
\caption{Constraints on $c_\infty^2$ set by each catalogue. Only two catalogues, HEAO and LRG, show a mildly peaked distribution at $c_\infty^2\simeq0.001$. The solid line shows the likelihood marginalised over the nuisance bias parameter, whereas the dashed line is obtained fixing the bias of each catalogue to its nominal value.}
\label{fig:cat_like}
\end{center}
\end{figure}

In order to investigate whether this peak is a signature of any peculiar process happening at a specific redshift we performed the analysis for each catalogue separately. The result is plotted in Fig.~\ref{fig:cat_like}. For many surveys the likelihood results monotonically decreasing with the speed of sound, peaking at the value which recovers the standard model. Two of them show actually a peak at $c^2_\infty=10^{-4}$: HEAO and LRG. By looking at the redshift distribution of the sources, Fig.~\ref{fig:ccf}, no common feature emerges as a clear indication of some physical process. The two distributions peak roughly at $z\simeq 0.6$, but the LRG one is extremely sharp, whereas HEAO very broad.

We conclude then that current galaxy surveys are not very sensitive to the speed of sound of the class of models we studied, the main reason being a not accurate enough redshift characterisation of the samples, which would allow us to perform a tomographic study of the time evolution of the cross-correlation function.

\subsection{Tomographic analysis}
\begin{figure}[h]
\begin{center}
\includegraphics[width = 0.49\textwidth]{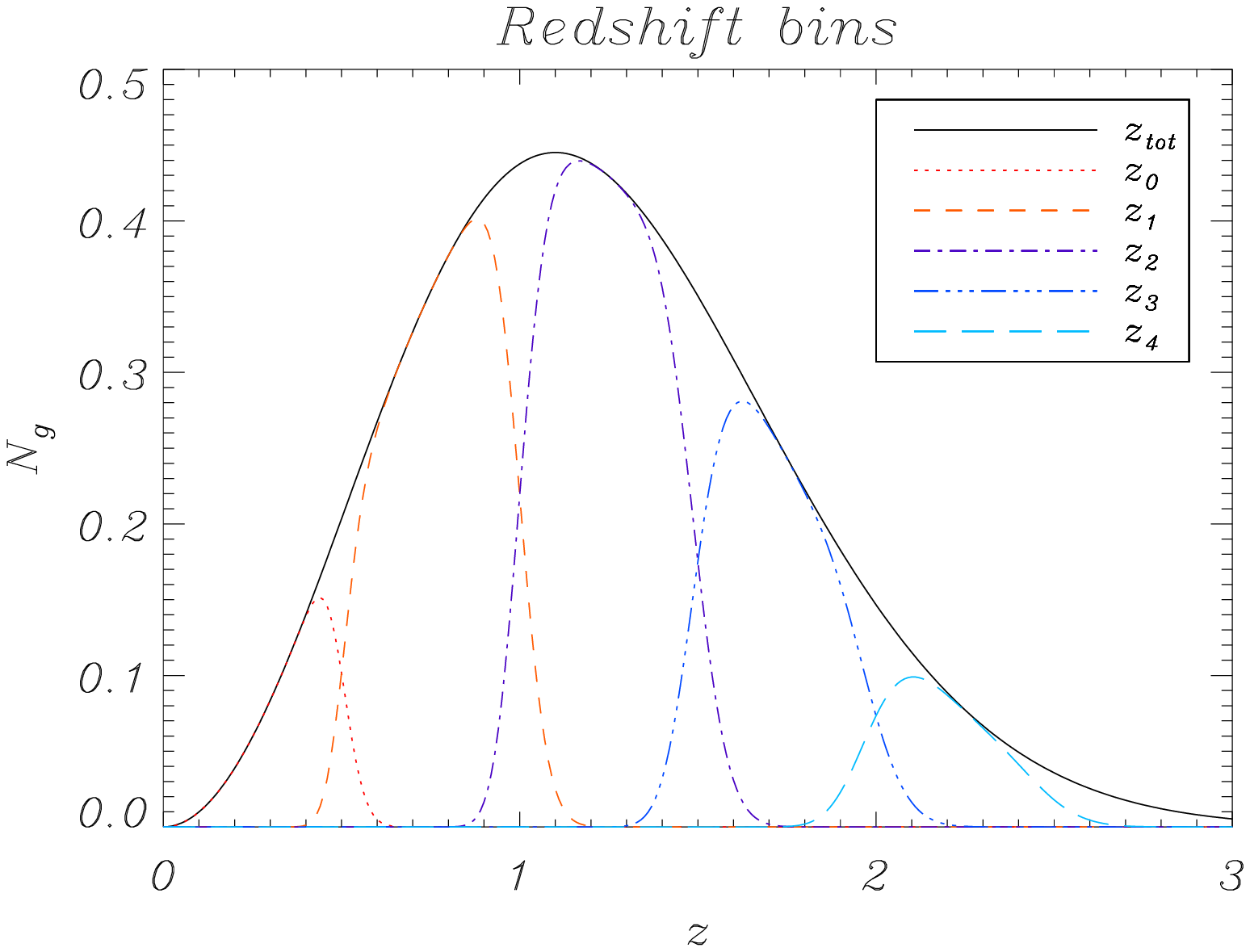}
\includegraphics[width = 0.49\textwidth]{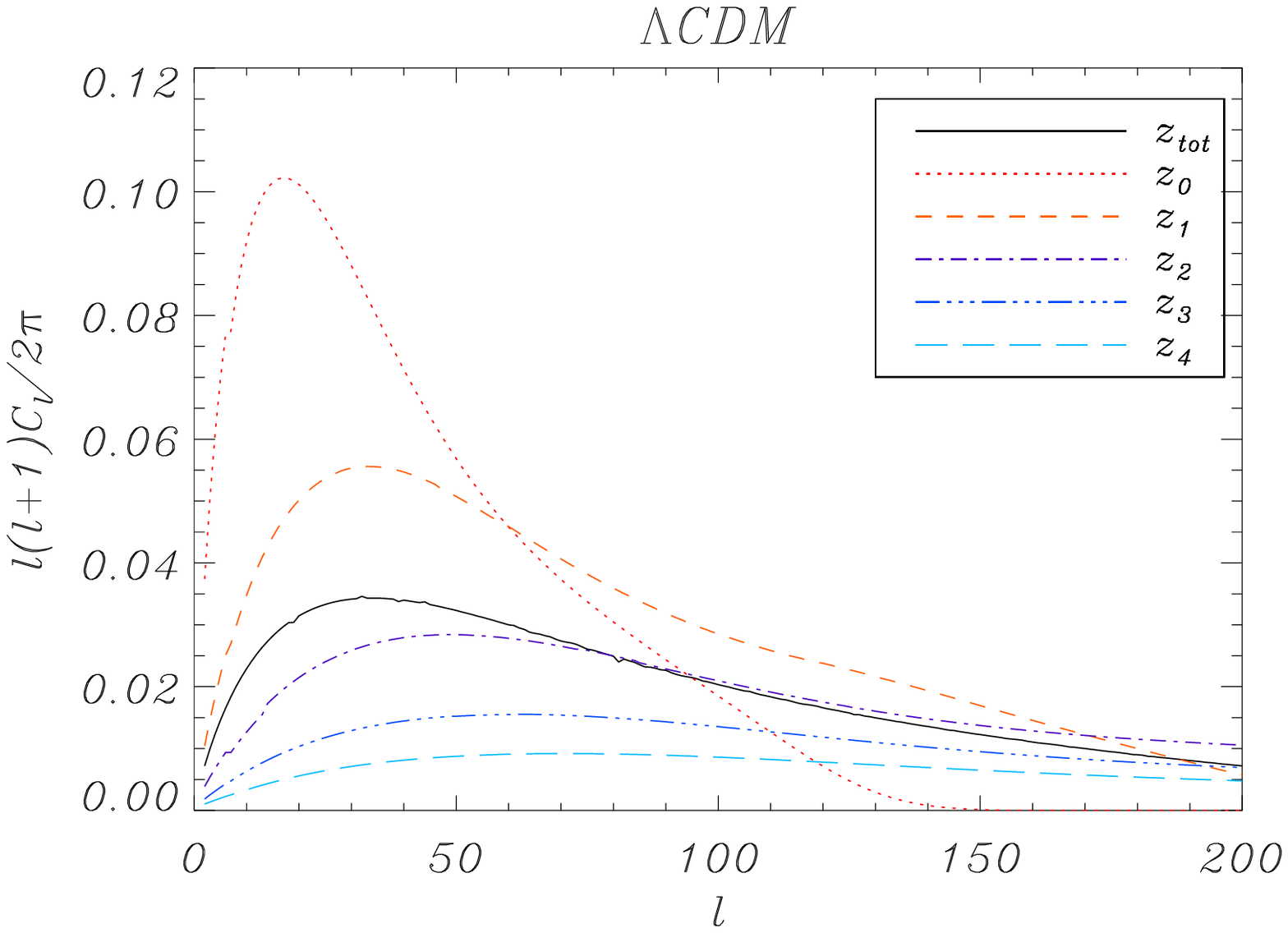}
\includegraphics[width = 0.49\textwidth]{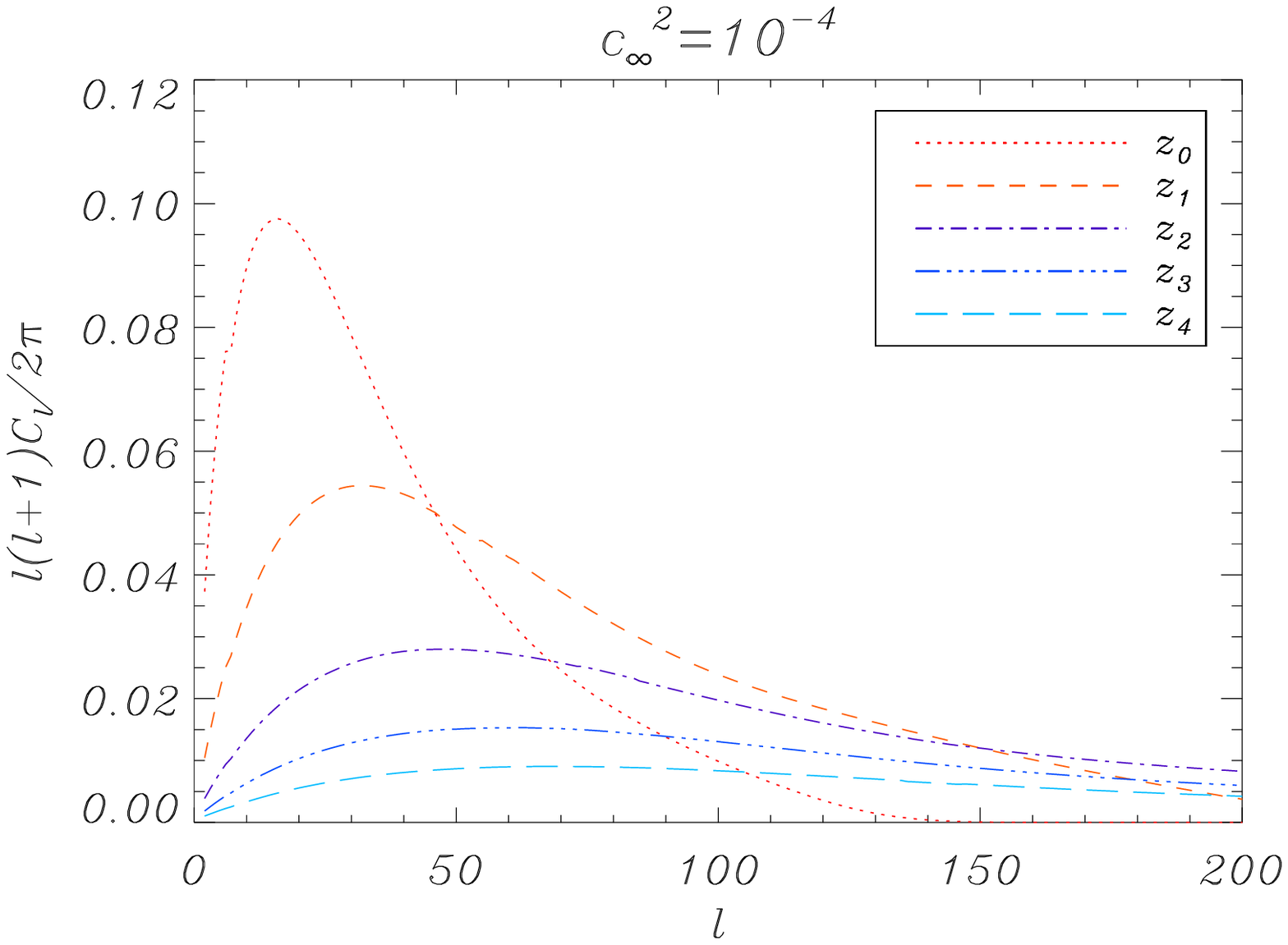}
\includegraphics[width = 0.49\textwidth]{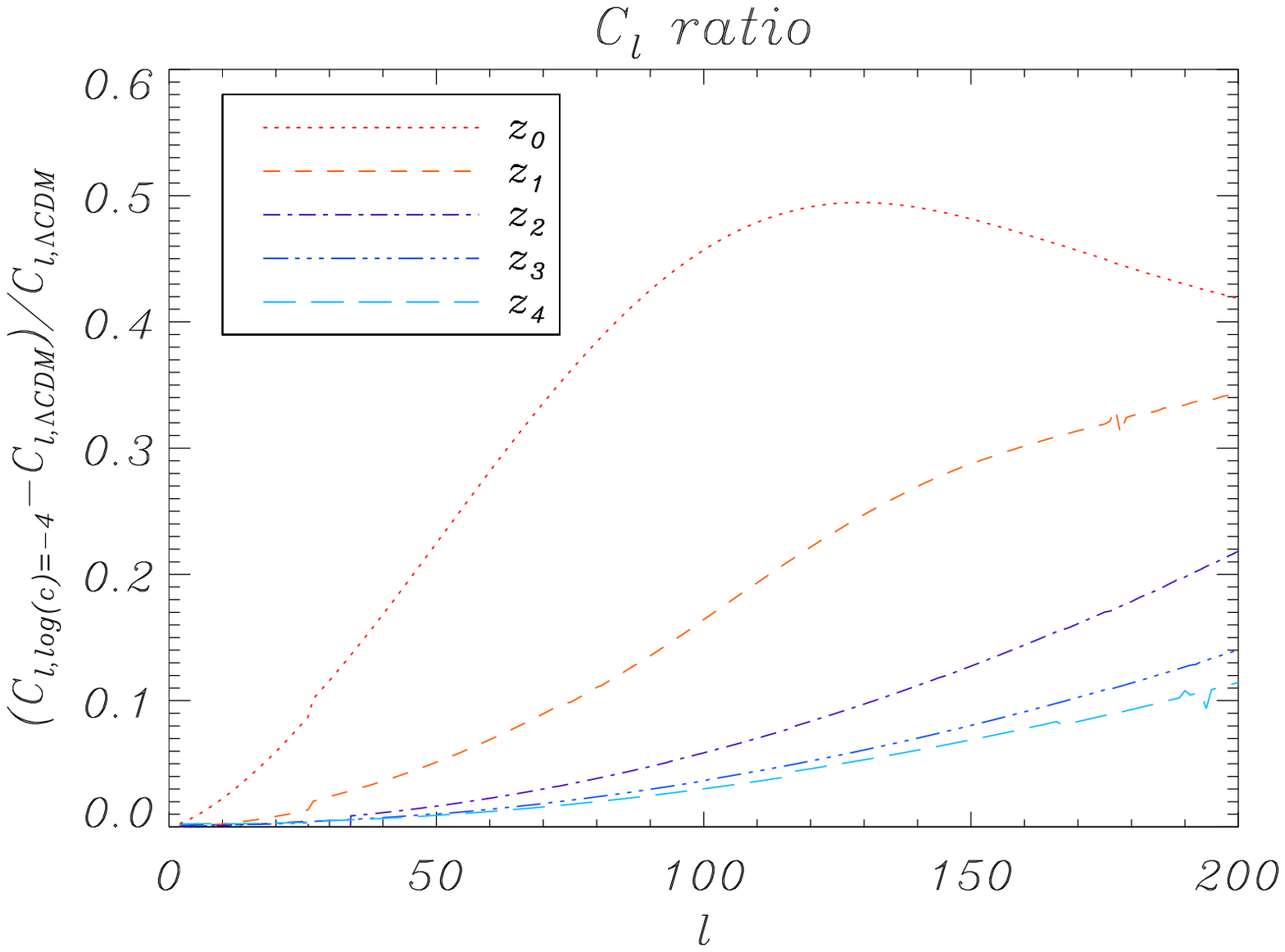}
\caption{{\emph{Top-left}}: Mock redshift distribution and redshift bins used to analyse the redshift dependence of ISW. {\emph{Top-right}}, \emph{bottom-left} and \emph{bottom-right}: angular (cross-)power spectrum for each redshift bin for the $\Lambda$CDM model, the choice $c_\infty^2=10^{-4}$ and the relative difference. As expected, the most sensitive bin is the first one, peaked at $z\sim0.4$; interestingly the largest difference between the models occurs at $\ell\sim100$.}
\label{fig:tomo}
\end{center}
\end{figure}

In order to test for which range in redshift we could have the best opportunity to discriminate between UDM models and $\Lambda$CDM, we perform a tomographic analysis, using five mock redshift distributions, following the model in Ref.\ \cite{Hu:2004yd} (see the left panel of Fig. \ref{fig:tomo}).

Since the standard model coincides with the choice $c_{\infty}^{2}=0$, with $c_{\infty}^{2}=10^{-7}$ being already a good approximation, and a UDM model with $c_{\infty}^{2}=10^{-4}$ seems to be slightly favoured by current observations, we compare the two cross-correlation functions computed for the same redshift slices as a function of the peak of the galaxy distribution, $z_{\rm p}$, looking for a signature of the different time evolution of the two models. In the right panel of Fig.~\ref{fig:tomo} are our results; as one can see, and as we expected (see Ref.~\cite{Camera:2009uz}), the maximum of the sensibility is for low-redshift surveys, so future tests of this model should be performed with tracers at low redshift.
Indeed, sources at lower $z_{\rm p}$ emit light that strongly feels the decay and the oscillations of the Newtonian potential, because this potential is sensitive to $k_{\rm J}(z,c_\infty)$, that decreases with time, and to the presence of an effective $\Omega_\Lambda$, that plays the role of DE \cite{Camera:2009uz, Bertacca:2007cv}. 


\section{Conclusions}
\label{conclusions}
In this work we have considered the cosmological models of UDM introduced in \cite{Bertacca:2008uf} focusing on predictions for the two-point correlation function between CMB and galaxy distribution. The main goal of this work is to investigate the ISW effect within the framework of UDM models based on a scalar field and understand which constraints we can put on the sound speed of the scalar field, crucial parameter of these models, using the CMB-LSS correlation. Analysing the cross-correlation between different source catalogs (NVSS, SDSS, HEAO, 2MASS) and the WMAP CMB temperature map as discussed in \cite{Giannantonio:2008}, we studied its changes due to a non-negligible sound speed and compared them to the standard  $\Lambda$CDM scenario. 

As expected, we found that for a sound speed smaller than $10^{-4}$, the UDM model gives the same prediction of the $\Lambda$CDM one, whilst for $c_{\infty}^2 > 10^{-4}$ the model starts to behave in a different way.
In particular, this effect can be explained through the oscillatory behaviour of the density contrast and of the gravitational potential when $k \gtrsim k_{\rm J}$ (see Eqs.~(\ref{Jeans}), (\ref{u-cs_k>>theta''/theta}) and (\ref{delta-dm})). Indeed this is due to two factors: {\it i}) through the amplitude of the density contrast that decays over time: the presence of a non negligible speed of sound prevents the structure formation on scale smaller than the Jeans' length (see Eq.~\ref{Jeans}), so that one expects a cross-correlation in these cases lower than in the $\Lambda $CDM model;
{\it ii}) the rapid fluctuations between positive and negative values produce a temporal mismatch between $\Phi'$ and $\delta_{\rm DM}$ which determines a lowering of the cross-correlation signal.


In conclusion, the comparison of theoretical predictions with real data has shown that current surveys are not precise enough to decisively distinguish between specific classes of late-time evolution of the Universe. Future surveys will have a better redshift characterisation of sources so that they will allow a tomographic study. We investigated this prospective creating a mock catalogue with a realistic redshift distribution. In particular, from tomographic analysis, the maximum of the sensibility is for low-redshift surveys. Indeed, sources at lower $z_{\rm p}$ emit light that strongly feels the decay and the oscillations of the Newtonian potential, because this potential is sensitive to $k_{\rm J}(z,c_\infty)$, that decreases with time, and to the presence of an effective $\Omega_\Lambda$, that plays the role of DE.

\vspace{0.5in}

\acknowledgments{DB would like to acknowledge the ICG at the University of Portsmouth for the hospitality during the development of this project and La ``Fondazione Ing. Aldo Gini" for support. DB research has been partly supported by ASI contract I/016/07/0 ``COFIS". AR is grateful for the support from a UK Science and Technology Facilities Research Council (STFC) PhD studentship. OFP research has been supported by the CNPq contract 150143/2010-9. Part of the research of DP was carried out at the Jet Propulsion Laboratory, California Institute of Technology, under a contract with the National Aeronautics and Space Administration. TG acknowledges support from the Alexander von Humboldt Foundation. The authors also thank  R.\ Crittenden, L.\ Verde, M.\ Viel for discussions and suggestions.}

\bibliographystyle{JHEP}
\bibliography{BCrossISW-LSS}

\providecommand{\href}[2]{#2}\begingroup\raggedright\begin{thebibliography}{10}

\bibitem{2mass}
T.~H. {Jarrett}, T.~{Chester}, R.~{Cutri}, S.~{Schneider}, M.~{Skrutskie}, and
  J.~P. {Huchra}, {\it {2MASS Extended Source Catalog: Overview and
  Algorithms}},  {\em \aj} {\bf 119} (May, 2000) 2498--2531,
  [\href{http://arxiv.org/abs/arXiv:astro-ph/0004318}{{\tt
  arXiv:astro-ph/0004318}}].

\bibitem{sdss_dr6}
{Adelman-McCarthy, J.~K., et al.}, {\it {The Sixth Data Release of the Sloan
  Digital Sky Survey}},  {\em \apjs} {\bf 175} (Apr., 2008) 297--313,
  [\href{http://arxiv.org/abs/0707.3413}{{\tt arXiv:0707.3413}}].

\bibitem{Condon1998NVSS}
J.~J. {Condon}, W.~D. {Cotton}, E.~W. {Greisen}, Q.~F. {Yin}, R.~A. {Perley},
  G.~B. {Taylor}, and J.~J. {Broderick}, {\it {The NRAO VLA Sky Survey}},  {\em
  \apj} {\bf 115} (May, 1998) 1693--1716.

\bibitem{heao}
E.~A. Boldt, {\it {THE COSMIC X-RAY BACKGROUND}},  {\em Phys. Rept.} {\bf 146}
  (1987) 215.

\bibitem{Perlmutter:1998np}
{\bf Supernova Cosmology Project} Collaboration, S.~Perlmutter {\em et~al.},
  {\it {Measurements of Omega and Lambda from 42 High-Redshift Supernovae}},
  {\em Astrophys. J.} {\bf 517} (1999) 565--586,
  [\href{http://arxiv.org/abs/astro-ph/9812133}{{\tt astro-ph/9812133}}].

\bibitem{Riess:1998cb}
{\bf Supernova Search Team} Collaboration, A.~G. Riess {\em et~al.}, {\it
  {Observational Evidence from Supernovae for an Accelerating Universe and a
  Cosmological Constant}},  {\em Astron. J.} {\bf 116} (1998) 1009--1038,
  [\href{http://arxiv.org/abs/astro-ph/9805201}{{\tt astro-ph/9805201}}].

\bibitem{Riess:1998dv}
A.~G. Riess {\em et~al.}, {\it {BVRI Light Curves for 22 Type Ia Supernovae}},
  {\em Astron. J.} {\bf 117} (1999) 707--724,
  [\href{http://arxiv.org/abs/astro-ph/9810291}{{\tt astro-ph/9810291}}].

\bibitem{Amanullah:2010vv}
R.~Amanullah {\em et~al.}, {\it {Spectra and Light Curves of Six Type Ia
  Supernovae at $0.511 < z < 1.12$ and the Union2 Compilation}},  {\em
  Astrophys. J.} {\bf 716} (2010) 712--738,
  [\href{http://arxiv.org/abs/1004.1711}{{\tt arXiv:1004.1711}}].

\bibitem{Larson:2010gs}
D.~Larson {\em et~al.}, {\it {Seven-Year Wilkinson Microwave Anisotropy Probe
  (WMAP) Observations: Power Spectra and WMAP-Derived Parameters}},
  \href{http://arxiv.org/abs/1001.4635}{{\tt arXiv:1001.4635}}.

\bibitem{Komatsu:2010fb}
E.~Komatsu {\em et~al.}, {\it {Seven-Year Wilkinson Microwave Anisotropy Probe
  (WMAP) Observations: Cosmological Interpretation}},
  \href{http://arxiv.org/abs/1001.4538}{{\tt arXiv:1001.4538}}.

\bibitem{Tsujikawa:2010sc}
S.~Tsujikawa, {\it {Dark energy: investigation and modeling}},
  \href{http://arxiv.org/abs/1004.1493}{{\tt arXiv:1004.1493}}.

\bibitem{Copeland:2006wr}
E.~J. Copeland, M.~Sami, and S.~Tsujikawa, {\it {Dynamics of dark energy}},
  {\em Int. J. Mod. Phys.} {\bf D15} (2006) 1753--1936,
  [\href{http://arxiv.org/abs/hep-th/0603057}{{\tt hep-th/0603057}}].

\bibitem{Adriani:2008zr}
{\bf PAMELA} Collaboration, O.~Adriani {\em et~al.}, {\it {An anomalous
  positron abundance in cosmic rays with energies 1.5-100 GeV}},  {\em Nature}
  {\bf 458} (2009) 607--609, [\href{http://arxiv.org/abs/0810.4995}{{\tt
  arXiv:0810.4995}}].

\bibitem{Adriani:2008zq}
O.~Adriani {\em et~al.}, {\it {A new measurement of the antiproton-to-proton
  flux ratio up to 100 GeV in the cosmic radiation}},  {\em Phys. Rev. Lett.}
  {\bf 102} (2009) 051101, [\href{http://arxiv.org/abs/0810.4994}{{\tt
  arXiv:0810.4994}}].

\bibitem{Bernabei:2000qi}
{\bf DAMA} Collaboration, R.~Bernabei {\em et~al.}, {\it {Search for WIMP
  annual modulation signature: Results from DAMA / NaI-3 and DAMA / NaI-4 and
  the global combined analysis}},  {\em Phys. Lett.} {\bf B480} (2000) 23--31.

\bibitem{Weinberg:1988cp}
S.~Weinberg, {\it {The cosmological constant problem}},  {\em Rev. Mod. Phys.}
  {\bf 61} (1989) 1--23.

\bibitem{Amendola:1272934}
L.~Amendola and S.~Tsujikawa, {\em Dark energy: Theory and observations}.
\newblock Cambridge Univ. Press, Cambridge, 2010.

\bibitem{Sachs:1967er}
R.~K. Sachs and A.~M. Wolfe, {\it {Perturbations of a cosmological model and
  angular variations of the microwave background}},  {\em Astrophys. J.} {\bf
  147} (1967) 73--90.

\bibitem{Crittenden:1995ak}
R.~G. Crittenden and N.~Turok, {\it {Looking for $\Lambda$ with the Rees-Sciama
  Effect}},  {\em Phys. Rev. Lett.} {\bf 76} (1996) 575,
  [\href{http://arxiv.org/abs/astro-ph/9510072}{{\tt astro-ph/9510072}}].

\bibitem{Boughn:2001zs}
S.~P. Boughn and R.~G. Crittenden, {\it {Cross-Correlation of the Cosmic
  Microwave Background with Radio Sources: Constraints on an Accelerating
  Universe}},  {\em Phys. Rev. Lett.} {\bf 88} (2002) 021302,
  [\href{http://arxiv.org/abs/astro-ph/0111281}{{\tt astro-ph/0111281}}].

\bibitem{Boughn:2003yz}
S.~Boughn and R.~Crittenden, {\it {A correlation of the cosmic microwave sky
  with large scale structure}},  {\em Nature} {\bf 427} (2004) 45--47,
  [\href{http://arxiv.org/abs/astro-ph/0305001}{{\tt astro-ph/0305001}}].

\bibitem{Jarosik:2010}
N.~{Jarosik}, C.~L. {Bennett}, J.~{Dunkley}, B.~{Gold}, M.~R. {Greason},
  M.~{Halpern}, R.~S. {Hill}, G.~{Hinshaw}, A.~{Kogut}, E.~{Komatsu},
  D.~{Larson}, M.~{Limon}, S.~S. {Meyer}, M.~R. {Nolta}, N.~{Odegard},
  L.~{Page}, K.~M. {Smith}, D.~N. {Spergel}, G.~S. {Tucker}, J.~L. {Weiland},
  E.~{Wollack}, and E.~L. {Wright}, {\it {Seven-Year Wilkinson Microwave
  Anisotropy Probe (WMAP) Observations: Sky Maps, Systematic Errors, and Basic
  Results}},  {\em ArXiv e-prints} (Jan., 2010)
  [\href{http://arxiv.org/abs/1001.4744}{{\tt arXiv:1001.4744}}].

\bibitem{Boldt:1987}
E.~{Boldt}, {\it {The cosmic X-ray background.}},  {\em \physrep} {\bf 146}
  (1987) 215--257.

\bibitem{Nolta:2003uy}
{\bf WMAP} Collaboration, M.~R. Nolta {\em et~al.}, {\it {First Year Wilkinson
  Microwave Anisotropy Probe (WMAP) Observations: Dark Energy Induced
  Correlation with Radio Sources}},  {\em Astrophys. J.} {\bf 608} (2004)
  10--15, [\href{http://arxiv.org/abs/astro-ph/0305097}{{\tt
  astro-ph/0305097}}].

\bibitem{Vielva:2006}
P.~{Vielva}, E.~{Mart{\'{\i}}nez-Gonz{\'a}lez}, and M.~{Tucci}, {\it
  {Cross-correlation of the cosmic microwave background and radio galaxies in
  real, harmonic and wavelet spaces: detection of the integrated Sachs-Wolfe
  effect and dark energy constraints}},  {\em \mnras} {\bf 365} (Jan., 2006)
  891--901, [\href{http://arxiv.org/abs/arXiv:astro-ph/0408252}{{\tt
  arXiv:astro-ph/0408252}}].

\bibitem{Cabre:2006qm}
A.~Cabre, E.~Gaztanaga, M.~Manera, P.~Fosalba, and F.~Castander, {\it
  {Cross-correlation of WMAP 3rd year and the SDSS DR4 galaxy survey: new
  evidence for Dark Energy}},  {\em Mon. Not. Roy. Astron. Soc.} {\bf 372}
  (2006) L23--L27, [\href{http://arxiv.org/abs/astro-ph/0603690}{{\tt
  astro-ph/0603690}}].

\bibitem{Gaztanaga:2004sk}
E.~Gaztanaga, M.~Manera, and T.~Multamaki, {\it {New light on dark cosmos}},
  {\em Mon. Not. Roy. Astron. Soc.} {\bf 365} (2006) 171--177,
  [\href{http://arxiv.org/abs/astro-ph/0407022}{{\tt astro-ph/0407022}}].

\bibitem{Giannantonio:2006}
T.~{Giannantonio}, R.~G. {Crittenden}, R.~C. {Nichol}, R.~{Scranton}, G.~T.
  {Richards}, A.~D. {Myers}, R.~J. {Brunner}, A.~G. {Gray}, A.~J. {Connolly},
  and D.~P. {Schneider}, {\it {High redshift detection of the integrated
  Sachs-Wolfe effect}},  {\em \prd} {\bf 74} (Sept., 2006) 063520--+,
  [\href{http://arxiv.org/abs/arXiv:astro-ph/0607572}{{\tt
  arXiv:astro-ph/0607572}}].

\bibitem{Pietrobon:2006}
D.~{Pietrobon}, A.~{Balbi}, and D.~{Marinucci}, {\it {Integrated Sachs-Wolfe
  effect from the cross correlation of WMAP 3year and the NRAO VLA sky survey
  data: New results and constraints on dark energy}},  {\em \prd} {\bf 74}
  (Aug., 2006) 043524--+,
  [\href{http://arxiv.org/abs/arXiv:astro-ph/0606475}{{\tt
  arXiv:astro-ph/0606475}}].

\bibitem{McEwen2007}
J.~D. {McEwen}, P.~{Vielva}, M.~P. {Hobson}, E.~{Mart{\'{\i}}nez-Gonz{\'a}lez},
  and A.~N. {Lasenby}, {\it {Detection of the integrated Sachs-Wolfe effect and
  corresponding dark energy constraints made with directional spherical
  wavelets}},  {\em \mnras} {\bf 376} (Apr., 2007) 1211--1226,
  [\href{http://arxiv.org/abs/arXiv:astro-ph/0602398}{{\tt
  arXiv:astro-ph/0602398}}].

\bibitem{Raccanelli:2008kx}
A.~Raccanelli, A.~Bonaldi, M.~Negrello, S.~Matarrese, G.~Tormen, and G.~D.
  Zotti, {\it A reassessment of the evidence of the integrated sachs-wolfe
  effect through the wmap-nvss correlation},  {\em Mon.Not.Roy.Astron.Soc.}
  {\bf 386} (2008) 2161, [\href{http://arxiv.org/abs/0802.0084}{{\tt
  arXiv:0802.0084}}].

\bibitem{Ho:2008bz}
S.~Ho, C.~Hirata, N.~Padmanabhan, U.~Seljak, and N.~Bahcall, {\it {Correlation
  of CMB with large-scale structure: I. ISW Tomography and Cosmological
  Implications}},  {\em \prd} {\bf 78} (2008) 043519,
  [\href{http://arxiv.org/abs/0801.0642}{{\tt arXiv:0801.0642}}].

\bibitem{Giannantonio:2008}
{Tommaso Giannantonio, Ryan Scranton, Robert G. Crittenden, Robert C. Nichol ,
  Stephen P. Boughn, Adam D. Myers, Gordon T. Richards }, {\it {Combined
  analysis of the integrated Sachs-Wolfe effect and cosmological
  implications}},  {\em \prd} {\bf 77} (2008) 123520,
  [\href{http://arxiv.org/abs/0801.4380}{{\tt arXiv:0801.4380}}].

\bibitem{Granett:2008ju}
B.~R. Granett, M.~C. Neyrinck, and I.~Szapudi, {\it {An Imprint of
  Super-Structures on the Microwave Background due to the Integrated
  Sachs-Wolfe Effect}},  {\em Astrophys. J.} {\bf 683} (aug, 2008) L99--L102,
  [\href{http://arxiv.org/abs/0805.3695}{{\tt arXiv:0805.3695}}].

\bibitem{Xia:2010pe}
J.-Q. Xia {\em et~al.}, {\it {Constraining Primordial Non-Gaussianity with
  High-Redshift Probes}},  {\em JCAP} {\bf 1008} (2010) 013,
  [\href{http://arxiv.org/abs/1007.1969}{{\tt arXiv:1007.1969}}].

\bibitem{Aghanim:2007bt}
N.~Aghanim, S.~Majumdar, and J.~Silk, {\it {Secondary anisotropies of the
  CMB}},  {\em Rept. Prog. Phys.} {\bf 71} (2008) 066902,
  [\href{http://arxiv.org/abs/0711.0518}{{\tt arXiv:0711.0518}}].

\bibitem{Hu:2004yd}
W.~Hu and R.~Scranton, {\it {Measuring Dark Energy Clustering with CMB-Galaxy
  Correlations}},  {\em Phys. Rev.} {\bf D70} (2004) 123002,
  [\href{http://arxiv.org/abs/astro-ph/0408456}{{\tt astro-ph/0408456}}].

\bibitem{Corasaniti:2005pq}
P.-S. Corasaniti, T.~Giannantonio, and A.~Melchiorri, {\it {Constraining dark
  energy with cross-correlated CMB and Large Scale Structure data}},  {\em
  Phys. Rev.} {\bf D71} (2005) 123521,
  [\href{http://arxiv.org/abs/astro-ph/0504115}{{\tt astro-ph/0504115}}].

\bibitem{Giannantonio:2006ij}
T.~Giannantonio and A.~Melchiorri, {\it {Chaplygin gas in light of recent
  integrated Sachs-Wolfe effect data}},  {\em Class. Quant. Grav.} {\bf 23}
  (2006) 4125--4132, [\href{http://arxiv.org/abs/gr-qc/0606030}{{\tt
  gr-qc/0606030}}].

\bibitem{Xia:2009}
J.~{Xia}, {\it {Constraint on coupled dark energy models from observations}},
  {\em \prd} {\bf 80} (Nov., 2009) 103514--+,
  [\href{http://arxiv.org/abs/0911.4820}{{\tt arXiv:0911.4820}}].

\bibitem{Xia:2009dr}
J.-Q. Xia, M.~Viel, C.~Baccigalupi, and S.~Matarrese, {\it {The High Redshift
  Integrated Sachs-Wolfe Effect}},  {\em JCAP} {\bf 0909} (2009) 003,
  [\href{http://arxiv.org/abs/0907.4753}{{\tt arXiv:0907.4753}}].

\bibitem{Massardi:2010sx}
M.~Massardi {\em et~al.}, {\it {A model for the cosmological evolution of low
  frequency radio sources}},  {\em Mon. Not. Roy. Astron. Soc.} {\bf 404}
  (2010) 532, [\href{http://arxiv.org/abs/1001.1069}{{\tt arXiv:1001.1069}}].

\bibitem{Giannantonio:2008fk}
T.~Giannantonio, Y.-S. Song, and K.~Koyama, {\it Detectability of a
  phantom-like braneworld model with the integrated sachs-wolfe effect},
  \href{http://arxiv.org/abs/0803.2238v1}{{\tt arXiv:0803.2238v1}}.

\bibitem{Nesseris:2010pc}
S.~Nesseris, A.~De~Felice, and S.~Tsujikawa, {\it {Observational constraints on
  Galileon cosmology}},  \href{http://arxiv.org/abs/1010.0407}{{\tt
  arXiv:1010.0407}}.

\bibitem{DeFelice:2010pv}
A.~De~Felice and S.~Tsujikawa, {\it {Cosmology of a covariant Galileon field}},
   {\em Phys. Rev. Lett.} {\bf 105} (2010) 111301,
  [\href{http://arxiv.org/abs/1007.2700}{{\tt arXiv:1007.2700}}].

\bibitem{Giannantonio:2007za}
T.~Giannantonio and R.~Crittenden, {\it {The effect of reionization on the
  CMB-density correlation}},  {\em Mon. Not. Roy. Astron. Soc.} {\bf 381}
  (2007) 819, [\href{http://arxiv.org/abs/0706.0274}{{\tt arXiv:0706.0274}}].

\bibitem{Kamenshchik:2001cp}
A.~Y. Kamenshchik, U.~Moschella, and V.~Pasquier, {\it {An alternative to
  quintessence}},  {\em Phys. Lett.} {\bf B511} (2001) 265--268,
  [\href{http://arxiv.org/abs/gr-qc/0103004}{{\tt gr-qc/0103004}}].

\bibitem{Bilic:2001cg}
N.~Bilic, G.~B. Tupper, and R.~D. Viollier, {\it {Unification of dark matter
  and dark energy: The inhomogeneous Chaplygin gas}},  {\em Phys. Lett.} {\bf
  B535} (2002) 17--21, [\href{http://arxiv.org/abs/astro-ph/0111325}{{\tt
  astro-ph/0111325}}].

\bibitem{Bento:2002ps}
M.~C. Bento, O.~Bertolami, and A.~A. Sen, {\it {Generalized Chaplygin gas,
  accelerated expansion and dark energy-matter unification}},  {\em Phys. Rev.}
  {\bf D66} (2002) 043507, [\href{http://arxiv.org/abs/gr-qc/0202064}{{\tt
  gr-qc/0202064}}].

\bibitem{Carturan:2002si}
D.~Carturan and F.~Finelli, {\it {Cosmological Effects of a Class of Fluid Dark
  Energy Models}},  {\em Phys. Rev.} {\bf D68} (2003) 103501,
  [\href{http://arxiv.org/abs/astro-ph/0211626}{{\tt astro-ph/0211626}}].

\bibitem{Amendola:2003bz}
L.~Amendola, F.~Finelli, C.~Burigana, and D.~Carturan, {\it {WMAP and the
  Generalized Chaplygin Gas}},  {\em JCAP} {\bf 0307} (2003) 005,
  [\href{http://arxiv.org/abs/astro-ph/0304325}{{\tt astro-ph/0304325}}].

\bibitem{Sandvik:2002jz}
H.~Sandvik, M.~Tegmark, M.~Zaldarriaga, and I.~Waga, {\it {The end of unified
  dark matter?}},  {\em Phys. Rev.} {\bf D69} (2004) 123524,
  [\href{http://arxiv.org/abs/astro-ph/0212114}{{\tt astro-ph/0212114}}].

\bibitem{Makler:2003iw}
M.~Makler, S.~Quinet~de Oliveira, and I.~Waga, {\it {Observational constraints
  on Chaplygin quartessence: Background results}},  {\em Phys. Rev.} {\bf D68}
  (2003) 123521, [\href{http://arxiv.org/abs/astro-ph/0306507}{{\tt
  astro-ph/0306507}}].

\bibitem{Scherrer:2004au}
R.~J. Scherrer, {\it {Purely kinetic k-essence as unified dark matter}},  {\em
  Phys. Rev. Lett.} {\bf 93} (2004) 011301,
  [\href{http://arxiv.org/abs/astro-ph/0402316}{{\tt astro-ph/0402316}}].

\bibitem{Giannakis:2005kr}
D.~Giannakis and W.~Hu, {\it {Kinetic unified dark matter}},  {\em Phys. Rev.}
  {\bf D72} (2005) 063502, [\href{http://arxiv.org/abs/astro-ph/0501423}{{\tt
  astro-ph/0501423}}].

\bibitem{Bertacca:2007ux}
D.~Bertacca, S.~Matarrese, and M.~Pietroni, {\it {Unified dark matter in scalar
  field cosmologies}},  {\em Mod. Phys. Lett.} {\bf A22} (2007) 2893--2907,
  [\href{http://arxiv.org/abs/astro-ph/0703259}{{\tt astro-ph/0703259}}].

\bibitem{Bertacca:2007cv}
D.~Bertacca and N.~Bartolo, {\it {ISW effect in Unified Dark Matter Scalar
  Field Cosmologies: an analytical approach}},  {\em JCAP} {\bf 0711} (2007)
  026, [\href{http://arxiv.org/abs/0707.4247}{{\tt arXiv:0707.4247}}].

\bibitem{Bertacca:2007fc}
D.~Bertacca, N.~Bartolo, and S.~Matarrese, {\it {Halos of Unified Dark Matter
  Scalar Field}},  {\em JCAP} {\bf 0805} (2008) 005,
  [\href{http://arxiv.org/abs/0712.0486}{{\tt arXiv:0712.0486}}].

\bibitem{Quercellini:2007ht}
C.~Quercellini, M.~Bruni, and A.~Balbi, {\it {Affine equation of state from
  quintessence and k-essence fields}},  {\em Class. Quant. Grav.} {\bf 24}
  (2007) 5413--5426, [\href{http://arxiv.org/abs/0706.3667}{{\tt
  arXiv:0706.3667}}].

\bibitem{Balbi:2007mz}
A.~Balbi, M.~Bruni, and C.~Quercellini, {\it {Lambda-alpha DM: Observational
  constraints on unified dark matter with constant speed of sound}},  {\em
  Phys. Rev.} {\bf D76} (2007) 103519,
  [\href{http://arxiv.org/abs/astro-ph/0702423}{{\tt astro-ph/0702423}}].

\bibitem{Bertacca:2008uf}
D.~Bertacca, N.~Bartolo, A.~Diaferio, and S.~Matarrese, {\it {How the Scalar
  Field of Unified Dark Matter Models Can Cluster}},  {\em JCAP} {\bf 0810}
  (2008) 023, [\href{http://arxiv.org/abs/0807.1020}{{\tt arXiv:0807.1020}}].

\bibitem{Pietrobon:2008js}
D.~Pietrobon, A.~Balbi, M.~Bruni, and C.~Quercellini, {\it {Affine
  parameterization of the dark sector: constraints from WMAP5 and SDSS}},  {\em
  Phys. Rev.} {\bf D78} (2008) 083510,
  [\href{http://arxiv.org/abs/0807.5077}{{\tt arXiv:0807.5077}}].

\bibitem{Bilic:2008yr}
N.~Bilic, G.~B. Tupper, and R.~D. Viollier, {\it {Cosmological tachyon
  condensation}},  {\em Phys. Rev.} {\bf D80} (2009) 023515,
  [\href{http://arxiv.org/abs/0809.0375}{{\tt arXiv:0809.0375}}].

\bibitem{Camera:2009uz}
S.~Camera, D.~Bertacca, A.~Diaferio, N.~Bartolo, and S.~Matarrese, {\it {Weak
  lensing signal in Unified Dark Matter models}},  {\em Mon. Not. Roy. Astron.
  Soc.} {\bf 399} (2009) 1995--2003,
  [\href{http://arxiv.org/abs/0902.4204}{{\tt arXiv:0902.4204}}].

\bibitem{Li:2009mf}
B.~Li and J.~D. Barrow, {\it {Does Bulk Viscosity Create a Viable Unified Dark
  Matter Model?}},  {\em Phys. Rev.} {\bf D79} (2009) 103521,
  [\href{http://arxiv.org/abs/0902.3163}{{\tt arXiv:0902.3163}}].

\bibitem{Chimento:2009nj}
L.~P. Chimento, R.~Lazkoz, and I.~Sendra, {\it {DBI models for the unification
  of dark matter and dark energy}},  {\em Gen. Rel. Grav.} {\bf 42} (2010)
  1189--1209, [\href{http://arxiv.org/abs/0904.1114}{{\tt arXiv:0904.1114}}].

\bibitem{Piattella:2009kt}
O.~F. Piattella, D.~Bertacca, M.~Bruni, and D.~Pietrobon, {\it {Unified Dark
  Matter models with fast transition}},  {\em JCAP} {\bf 1001} (2010) 014,
  [\href{http://arxiv.org/abs/0911.2664}{{\tt arXiv:0911.2664}}].

\bibitem{Gao:2009me}
C.~Gao, M.~Kunz, A.~R. Liddle, and D.~Parkinson, {\it {Unified dark energy and
  dark matter from a scalar field different from quintessence}},  {\em Phys.
  Rev.} {\bf D81} (2010) 043520, [\href{http://arxiv.org/abs/0912.0949}{{\tt
  arXiv:0912.0949}}].

\bibitem{Camera:2010wm}
S.~Camera, T.~D. Kitching, A.~F. Heavens, D.~Bertacca, and A.~Diaferio, {\it
  {Measuring Unified Dark Matter with 3D cosmic shear}},
  \href{http://arxiv.org/abs/1002.4740}{{\tt arXiv:1002.4740}}.

\bibitem{Lim:2010yk}
E.~A. Lim, I.~Sawicki, and A.~Vikman, {\it {Dust of Dark Energy}},  {\em JCAP}
  {\bf 1005} (2010) 012, [\href{http://arxiv.org/abs/1003.5751}{{\tt
  arXiv:1003.5751}}].

\bibitem{Bertacca-2010-2}
D.~{Bertacca}, M.~{Bruni}, O.~F. {Piattella}, and D.~{Pietrobon}, {\it {Unified
  Dark Matter scalar field models with fast transition}},  {\em ArXiv e-prints}
  (nov, 2010) [\href{http://arxiv.org/abs/1011.6669}{{\tt arXiv:1011.6669}}].

\bibitem{transfer-function}
O.~F. Piattella and D.~Bertacca, {\it {Gravitational potential evolution in
  Unified Dark Matter Scalar Field Cosmologies: an analytical approach}},
  \href{http://arxiv.org/abs/1103.0234}{{\tt arXiv:1103.0234}}.

\bibitem{Bertacca:2010ct}
D.~Bertacca, N.~Bartolo, and S.~Matarrese, {\it {Unified Dark Matter Scalar
  Field Models}},  {\em Advances in Astronomy} {\bf {2010}} ({2010})
  [\href{http://arxiv.org/abs/1008.0614}{{\tt arXiv:1008.0614}}].

\bibitem{Mukhanov:2005sc}
V.~Mukhanov, {\it {Physical foundations of cosmology}}, . Cambridge, UK: Univ.
  Pr. (2005) 421 p.

\bibitem{Mukhanov:1990me}
V.~F. Mukhanov, H.~A. Feldman, and R.~H. Brandenberger, {\it {Theory of
  cosmological perturbations. Part 1. Classical perturbations. Part 2. Quantum
  theory of perturbations. Part 3. Extensions}},  {\em Phys. Rept.} {\bf 215}
  (1992) 203--333.

\bibitem{Garriga:1999vw}
J.~Garriga and V.~F. Mukhanov, {\it {Perturbations in k-inflation}},  {\em
  Phys. Lett.} {\bf B458} (1999) 219--225,
  [\href{http://arxiv.org/abs/hep-th/9904176}{{\tt hep-th/9904176}}].

\bibitem{Hu:1998tj}
W.~Hu and D.~J. Eisenstein, {\it {The Structure of structure formation
  theories}},  {\em Phys. Rev.} {\bf D59} (1999) 083509,
  [\href{http://arxiv.org/abs/astro-ph/9809368}{{\tt astro-ph/9809368}}].

\bibitem{Dodelson:2003ft}
S.~Dodelson, {\it {Modern cosmology}}, . Amsterdam, Netherlands: Academic Pr.
  (2003) 440 p.

\bibitem{Eisenstein:1997ik}
D.~J. Eisenstein and W.~Hu, {\it {Baryonic Features in the Matter Transfer
  Function}},  {\em Astrophys. J.} {\bf 496} (1998) 605,
  [\href{http://arxiv.org/abs/astro-ph/9709112}{{\tt astro-ph/9709112}}].

\bibitem{Lahav:1991wc}
O.~Lahav, P.~B. Lilje, J.~R. Primack, and M.~J. Rees, {\it {Dynamical effects
  of the cosmological constant}},  {\em Mon. Not. Roy. Astron. Soc.} {\bf 251}
  (1991) 128--136.

\bibitem{Carroll:1991mt}
S.~M. Carroll, W.~H. Press, and E.~L. Turner, {\it {The Cosmological
  constant}},  {\em Ann. Rev. Astron. Astrophys.} {\bf 30} (1992) 499--542.

\bibitem{Eisenstein:1997ij}
D.~J. Eisenstein, {\it {An Analytic Expression for the Growth Function in a
  Flat Universe with a Cosmological Constant}},
  \href{http://arxiv.org/abs/astro-ph/9709054}{{\tt astro-ph/9709054}}.

\bibitem{Dunkley:2009}
J.~{Dunkley}, E.~{Komatsu}, M.~R. {Nolta}, D.~N. {Spergel}, D.~{Larson},
  G.~{Hinshaw}, L.~{Page}, C.~L. {Bennett}, B.~{Gold}, N.~{Jarosik}, J.~L.
  {Weiland}, M.~{Halpern}, R.~S. {Hill}, A.~{Kogut}, M.~{Limon}, S.~S. {Meyer},
  G.~S. {Tucker}, E.~{Wollack}, and E.~L. {Wright}, {\it {Five-Year Wilkinson
  Microwave Anisotropy Probe Observations: Likelihoods and Parameters from the
  WMAP Data}},  {\em \apjs} {\bf 180} (Feb., 2009) 306--329,
  [\href{http://arxiv.org/abs/0803.0586}{{\tt arXiv:0803.0586}}].

\end{thebibliography}\endgroup

\end{document}